\documentclass[5p]{elsarticle}
\usepackage{lineno,hyperref}
\usepackage{bm}
\hypersetup{
colorlinks = true,
linkcolor = blue,
anchorcolor = blue,
citecolor = blue,
filecolor = blue,
urlcolor = blue
}
\usepackage{xcolor}
\usepackage{amsmath}
\usepackage{float}
\usepackage[colorinlistoftodos]{todonotes}

\usepackage[nameinlink,capitalize]{cleveref}
\usepackage{placeins}
\usepackage{caption}
\usepackage[normalem]{ulem}
\useunder{\uline}{\ul}{}
\journal{Mechanics of Materials}
\biboptions{authoryear}

\begin{document}
\sloppy
%\modulolinenumbers[5]

\begin{frontmatter}

%\title{Magnesium alloys strengthened by nanoscale precipitates and fine solute clusters }
%\title{Strengthening magnesium alloys with solute atoms, solute clusters, and nanoscale precipitates}
\title{Strengthening magnesium by design: integrating alloying and dynamic processing}

\author[1,2]{Suhas Eswarappa Prameela\corref{corrauthor}}
\ead{seswara3@jhu.edu}
%\fnref{myfootnote}}
\author[1,2]{Peng Yi}
\author[5]{Yannick Hollenweger}
\author[6]{Burigede Liu}
\author[1,2]{Joey Chen}
\author[2]{Laszlo Kecskes}
\author[5]{Dennis M. Kochmann}
\author[1,2,3,4]{Michael L. Falk}
\author[1,2]{Timothy P. Weihs}


\cortext[corrauthor]{Corresponding author}

\address[1]{Department of Materials Science and Engineering, Johns Hopkins University, Baltimore, MD 21218, USA}
\address[2]{Hopkins Extreme Materials Institute, Johns Hopkins University, Baltimore, MD, 21218, USA }
\address[3]{Department of Mechanical Engineering, Johns Hopkins University, Baltimore, MD 21218, USA}
\address[4]{Department of Physics and Astronomy, Johns Hopkins University, Baltimore, MD 21218, USA}
\address[5]{Mechanics \& Materials Lab, Department of Mechanical and Process Engineering, ETH Zürich, 8092 Zürich, Switzerland }
\address[6]{Division of Engineering and Applied Science, California Institute of Technology, Pasadena, CA 91125, USA}
\begin{abstract}
Magnesium (Mg) has the lowest density of all structural metals and has excellent potential for wide use in structural applications. While pure Mg has inferior mechanical properties; the addition of further elements at various concentrations has produced alloys with enhanced mechanical performance and corrosion resistance. An important consequence of adding such elements is that the saturated Mg matrix can locally decompose to form solute clusters and intermetallic particles, often referred to as precipitates. Controlling the shape, number density, volume fraction, and spatial distribution of solute clusters and precipitates significantly impacts the alloy's plastic response. Conversely, plastic deformation during thermomechanical processing can dramatically impact solute clustering and precipitation. In this paper, we first discuss how solute atoms, solute clusters, and precipitates can improve the mechanical properties of Mg alloys. We do so by primarily comparing three alloy systems: Mg-Al, Mg-Zn, and Mg-Y-based alloys. In the second part, we provide strategies for optimizing such microstructures by controlling nucleation and growth of solute clusters and precipitates during thermomechanical processing. In the third part, we briefly highlight how one can enable inverse design of Mg alloys by a more robust Integrated Computational Materials Design (ICMD) approach.
\end{abstract}

\begin{keyword}
Magnesium, Precipitation, Nucleation, Clustering, ICMD 
%\texttt{elsarticle.cls}\sep \LaTeX\sep Elsevier \sep template
%\MSC[2010] 00-01\sep  99-00
\end{keyword}

\end{frontmatter}
\tableofcontents
%\linenumbers
%add 'Nomenclature' right above
% \usepackage{color}
\captionsetup[figure]{labelfont={bf},labelformat={default},labelsep=period,name={Fig.}}

\begin{table*}
\centering
\begin{tabular}{|llll|} 
\hline
\multicolumn{4}{|l|}{\textbf{Terminology} } \\
&&&\\
$\theta$ & Angle & $\epsilon$ & Strain \\
$A$ & Area & $n$ & Strain hardening exponent \\
$n_s$ & Areal number density & $\dot{\epsilon}$ & Strain rate \\
$s$ & Atomic area on slip plane & $a$ & Swept area \\
$\omega$ & Atomic volume of the matrix & $M$ & Taylor factor' \\
$v_m$ & Average atomic volume of matrix & $T$ & Temperature \\
$v_p$ & Average atomic volume of   precipitate & $V$ & Volume \\
$\phi$ & Binding angle & $W$ & Work \\
$k_b$ & Boltzmann constant & $\Theta$ & Work hardening rate \\
$K$ & Bulk modulus & \multicolumn{2}{l|}{\textbf{Alloy Designations (All wt.\%)}} \\
$b$ & Burger’s vector &  &  \\
$\alpha$ & Compositional strain   co-efficient & A$n$ & $n$\% Al, rest Mg \\
$\tau_r$ & Critical resolved shear stress & Z$n$ & $n$\% Zn, rest Mg \\
$M$ & Diffusion potential & W$n$ & $n$\% Y, rest Mg \\
$r_0$ & Dislocation core radius & AA7075 & 6\% Zn, 2.3\% Mg, 1.4\% Cu, rest   Al \\
$\rho$ & Dislocation density & AA6061 & 1\%Mg, 0.6\% Si, 0.3\% Cu, rest   Al \\
$\psi$ & Dissipation potential & AZ91 & 9\% Al, 1\% Zn, rest Mg \\
$S$ & Entropy & HM21 & 2\% Th, 1\% Zn, rest Mg \\
$\dot{\varepsilon}_0$ & Fitting parameter & ZK60 & 6\% Zn, 0.5\% Zr, rest Mg \\
$G$ & Formation energy & W5-Z0.1 & 5\% Y, 0.1\% Zn, rest Mg \\
$g_v$ & Free energy of precipitate per   volume of matrix & W20-Z9 & 20\% Y, 9\% Zn, rest Mg \\
$g$ & Gibbs free energy & WE54 & 5\% Y, 3\% Nd, 0.6\% Zr, rest Mg \\
$d_s$ & Glide plane spacing of   precipitates & Z8-RE1.5 & 8\% Zn, 1.5\% Rare-earth mix,   rest Mg \\
$v$ & Glide velocity & WE43/W4-RE3 & 4\% Y, 3\% Rare-earth mix, rest   Mg \\
$Q$ & Heat & Z6W1.5K0.5 & 6\% Zn, 1.5\% Zr, 0.5\% Y, rest   Mg \\
$\mathcal{H}$ & Helmholtz Free Energy & \multicolumn{2}{l|}{\textbf{Abbreviations}} \\
$P$ & Hydrostatic pressure field & \textbf{} &  \\
$w$ & Interaction energy & APT & Atomic Probe Tomography \\
$f_m$ & Interaction strength & CE model & Cottrell-Eshelby Model \\
$\upsilon$ & Interfacial energy & CNGT & Classical Nucleation and Growth   Theory \\
$U$ & Internal energy & CNT & Classical Nucleation Theory \\
$\lambda_e$ & Inter-particle spacing & CRSS & Critical Resolved Shear Stress \\
$L$ & Length & DIP & Deformation   Induced   Precipitation \\
$T_L$ & Line tension & EBSD & Electron Backscatter Diffraction \\
$\Lambda$ & Mean free path & ECAE & Equal Channel Angular Extrusion \\
$g_s$ & Misfit elastic strain energy & GP zone & Guinier–Preston zone \\
$e$ & Misfit parameter & HAADF & High-angle annular dark-field \\
$\sigma$ & Normal stress & HCP & Hexagonal close-packed \\
$I$ & Nucleation rate & HPT & High pressure torsion \\
$r_p$ & Planar radius & ICMD & Integrated Computational   Materials Design \\
$\nu$ & Poisson's ratio & MC & Monte-Carlo \\
$\bar{\tau}$ & Positron lifetime & MD & Molecular Dynamics \\
$d_t$ & Precipitate diameter ($=2r_p$) & NMR & Nuclear Magnetic Resonance \\
$t_t$ & Precipitate thickness & PAS & Positron   Annihilation   Spectroscopy \\
$f$ & Precipitate volume Fraction & PTS & Predominant twin system \\
$r$ & Radial distance & RE & Rare Earth \\
$R$ & Radius & SAXS & Small Angle X-ray   Scattering \\
$\tau_s$ & Resolved shear stress & SSSS & Supersaturated Solid Solution \\
$G_m$ & Shear modulus & STEM & Scanning Transmission Electron   Microscopy \\
$\gamma$ & Shear strain & TD & Twin Dislocation\textbackslash{}Disconnection \\
$\tau$ & Shear stress & TEM & Transmission Electron Microscopy \\
$\Delta   E_b$ & Solute barrier height & XRD & X-ray Diffraction \\
$c$ & Solute concentration &  &  \\

\hline

\end{tabular}
\end{table*} 
\newpage 
%\textcolor{white}{w}

%\newpage
\section{Introduction}
\label{sec:Introduction}
Magnesium (Mg) is the lightest structural metal. It is abundant, making up $~$2.7\% of the Earth’s crust, readily available and producible at high purity on a large scale \citep{Nie2020SESFE}. Furthermore, it has a high specific strength, making it a great candidate for light weight applications, such as aerospace, defense, biomedical, and transportation. However, in its pure elemental state,  it is too weak to be used for practical structural applications. This issue is typically addressed by alloying the pure material with small fractions of other elements ($<$ 15 wt\%) to Mg. On average, Mg alloys are about two-thirds as dense as aluminum (Al) alloys and only a quarter as dense as steel. Industrial applications of Mg include aircraft fuselage frames, automobiles, light bicycles, and -- due to Mg's great biocompatibility-- bioabsorbable implants \citep{luo2002magnesium,mordike2001magnesium,zhao2017current,froes1998science,furuya2000applications}. These environments demand extreme performances of the materials, either due to their loading, spatial constraints, or relationship to surrounding components. \\

The advantages of manufacturing lighter structural parts from Mg are evident; reducing the weight of aircraft or rockets increases their range or payload; lighter cars correlate directly with reduced fuel consumption; lighter armor plating decreases the weight a soldier must carry in the field, thereby enhancing mobility. Additionally, Mg production consumes less energy than Al production, and Mg's abundance makes it a more reliable long-term resource.  \\

\begin{figure}[h]
    \centering
    \includegraphics[width=\linewidth]{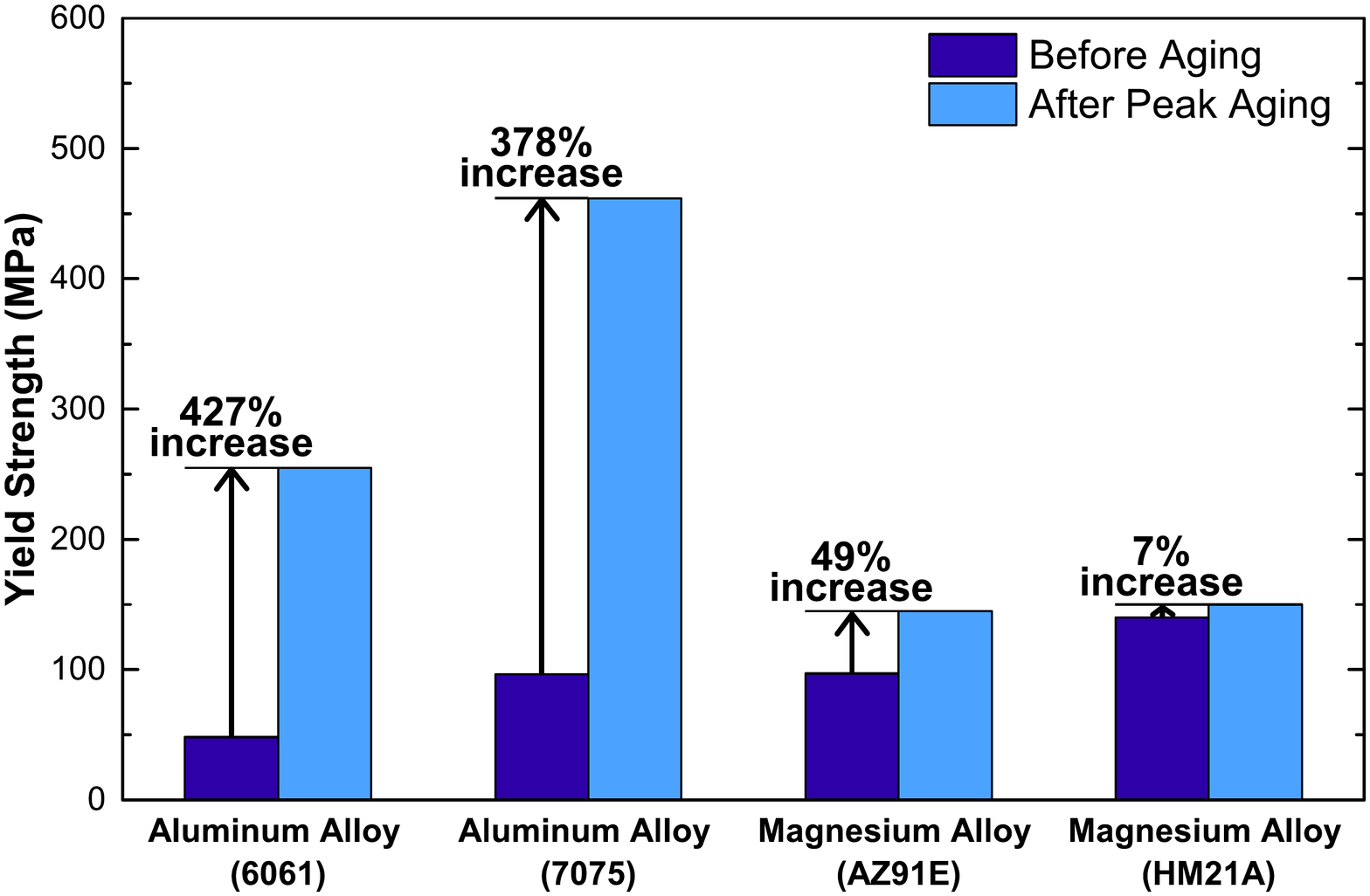}
    \caption{Comparison of pre- and post-heat treatment yield strengths of selected commercial Mg and Al alloys \citep{asm1990metals, bauccio1993asm, handbook1996edition, avedesian1999asm, alloys2015international, aluminum2000aluminum}. }
    \label{fig:AgingStrengths}
\end{figure}

However, many challenges remain in making Mg-based materials a viable alternative to Al and steels. In contrast to Al-based materials (and steels), Mg alloys show poor hardening with conventional thermal aging techniques, see \cref{fig:AgingStrengths}. \\

% \noindent Today's most prevalent low-density, high-strength metallic alloys across all industries are Al-based. Many have tried to mimic the strategies used in Al alloys for Mg alloys, with limited success. For instance, conventional heat treatment in Al alloys results in a 3-4 times increase in yield strength.
% Mg alloys, however, respond poorly to the same processes, as shown in \cref{fig:AgingStrengths}.

\begin{figure}[h]
    \centering
    \includegraphics[width=\linewidth]{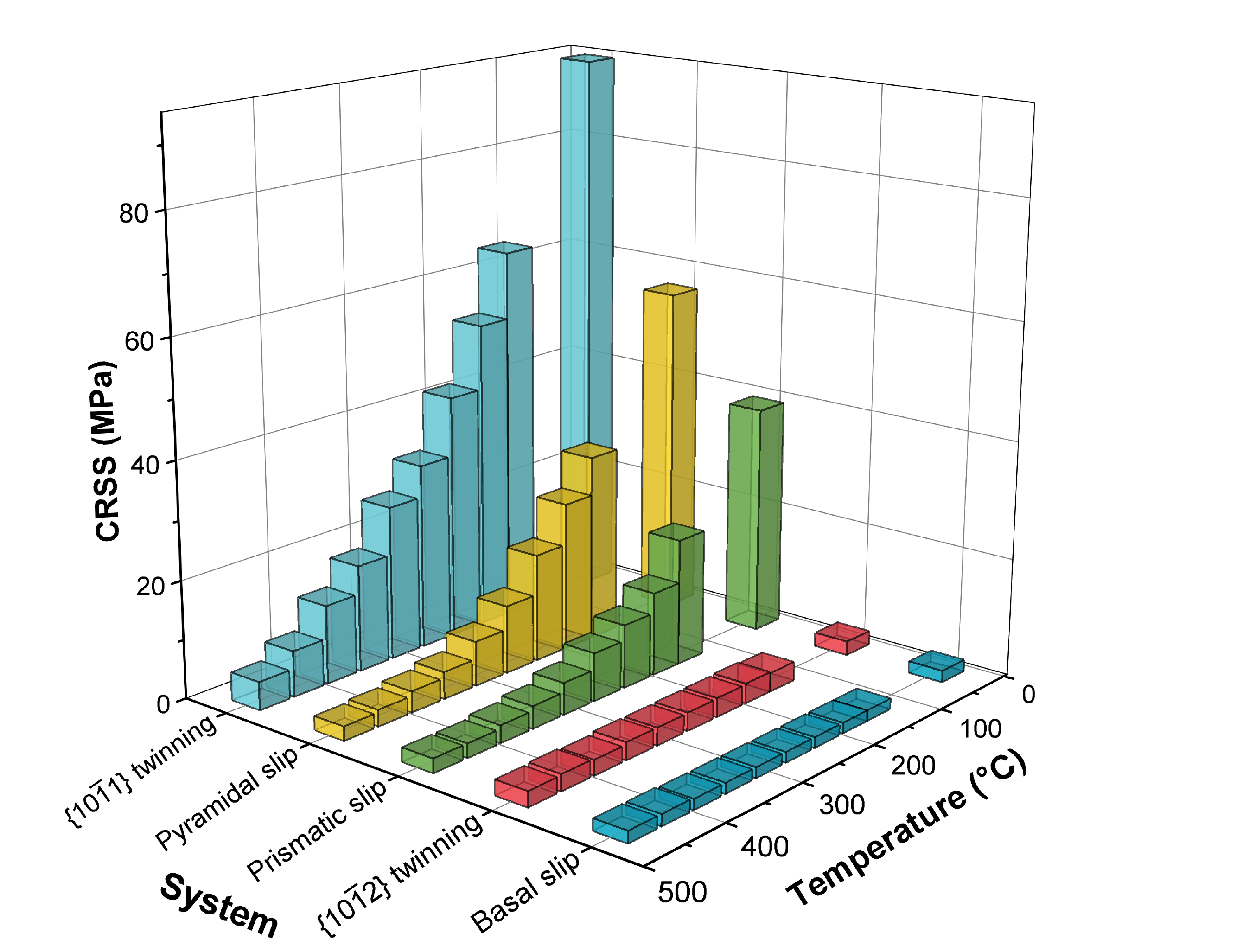}
    \caption{CRSS for slip and twinning systems for pure Mg single-crystals. Data adapted from \cite{nie2020microstructure}}
    \label{fig:CRSS3DBar}
\end{figure}

This discrepancy is in part due to fundamental differences in the crystallographic structure between Al and Mg. In comparison to the cubic crystal symmetry of Al, with 12 equivalent slip systems, Mg -- which shows a hexagonal close-packed symmetry (HCP) --  has four different slip variants: basal$\langle a \rangle$, prismatic$\langle a \rangle$, pyramidal$\langle a \rangle$, and pyramidal$\langle c+a \rangle$ (pyramidal I $\&$ II). These distinct slip systems have very dissimilar critical resolved shear stress (CRSS) values, which vary independently with temperature, as shown in \cref{fig:CRSS3DBar}, with basal slip being the easiest mode to activate. This induces a strong plastic anisotropy across a range of length scales and profoundly affects many mechanical properties and phase transformations.\\

\noindent Additionally, the lack of five independent slip systems and the inability of basal dislocations to accommodate arbitrary distortions results in Mg alloys not being able to fulfill the von Mises flow criterion \citep{mises1913mechanik}. This inability leads to an alternate deformation mechanism in Mg and its alloys, called twinning.\\

\begin{figure}[h]
    \centering
    \includegraphics[width=\linewidth]{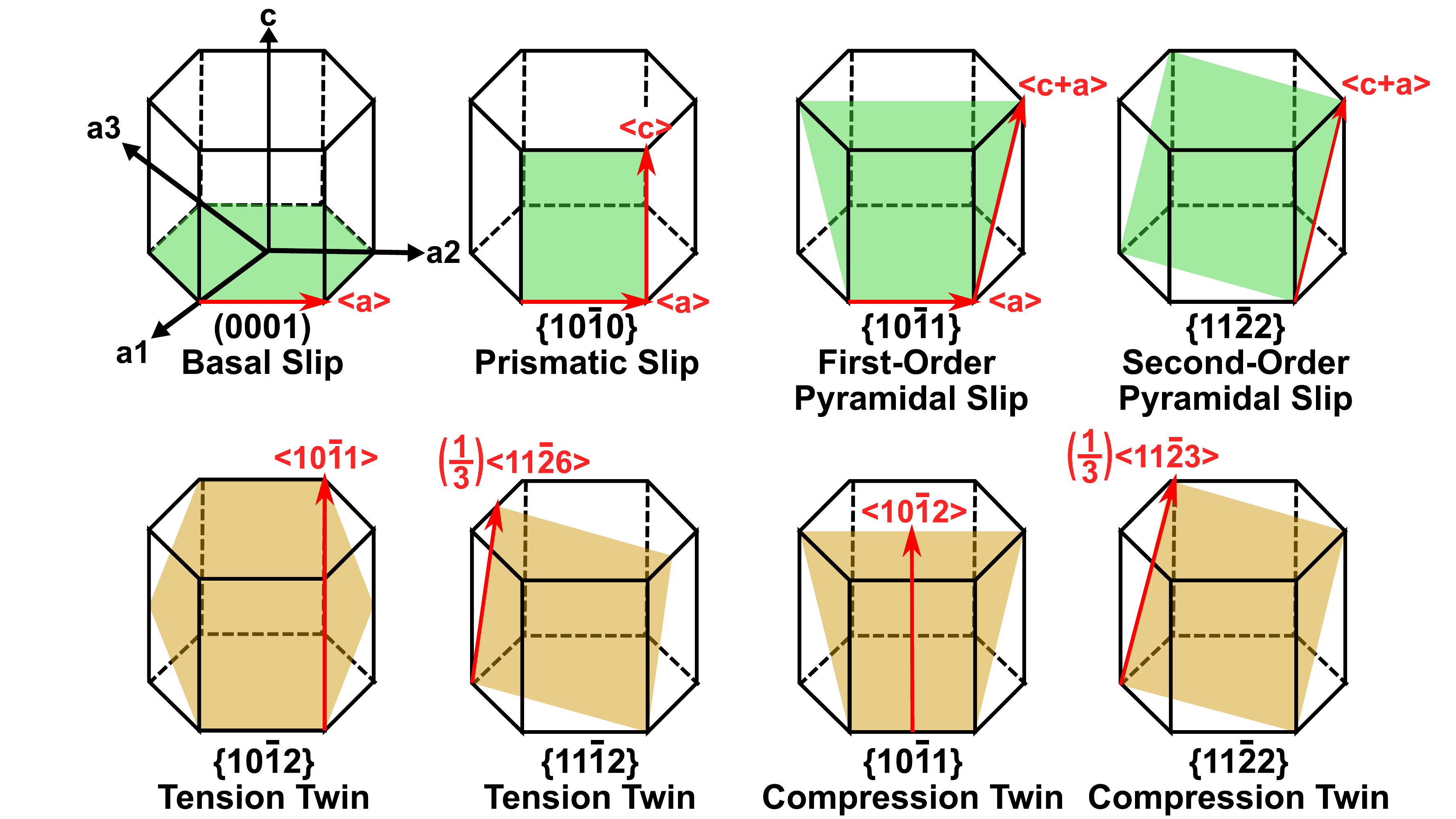}
    \caption{Slip and twinning mechanisms present in Mg. Figure adapted from \cite{fan2015molecular}}
    \label{fig:DeformationMechanisms}
\end{figure}

Twinning and slip (as shown in \cref{fig:DeformationMechanisms}) depend on parameters such as the $c/a$-ratio, the type of the alloying elements, the grain size,  the strain rate, and the temperature. Furthermore, the anisotropic nature of these plastic deformation mechanisms leads to a strong texture effect, and the hardening mechanisms affect different slip variants and twinning unequally. Hence, boosting strength and ductility of Mg alloys, while controlling twinning and limiting anisotropy, is a challenge that becomes more difficult in the context of designing alloys for applications in extreme environments (\cite{mallick2020brief, mallick2019laser, williams2020concise}).  \\

%The above models describe plasticity in Mg at the single-crystal level and hence require the simulation of fully-resolved representative polycrystalline unit cells at the mesoscale \citep{TutcuogluEtAl2019} or the use of Taylor-type models that provide an upper-bound response by averaging over a set of grains for a given orientation distribution function. The latter is another efficient reduced-order approach at the polycrystal level, which has been used successfully, e.g., in multiscale simulations of metal processing, such as equal channel angular extrusion \citep{kumar2020meshless}. Such approaches also aim to account for recrystallization while remaining computationally tractable, e.g. see \citep{kumar2020meshless}.\\ 

This paper focuses on strengthening binary Mg alloys by using mechanical processing to optimize their microstructures. Typically, two inter-related strategies are employed to design Mg alloys for improved mechanical performance. One alloys Mg with different elements at varying concentrations, while the other subjects the alloys to different techniques and routes of thermomechanical processing \citep{prameela2020binary}. Both approaches are geared towards manipulating the $c/a$-ratio, the grain size, the texture, and the CRSS values for dislocation glide, along with the characteristics of precipitates (chemistry, habit planes, size, and density). These efforts seek to induce particular strengthening mechanisms, namely grain size strengthening, precipitation strengthening, and solute strengthening. \\

The current work focuses on the use of solute atoms, solute clusters, and precipitates to boost a material's resistance to plastic deformation and, more importantly, on our ability to form fine clusters and nanoscale precipitates in Mg alloys by leveraging mechanical deformation. Consider a Mg alloy which was fully solutionized through a thermal process at an elevated temperature. Quenching, followed by aging of this alloy, promotes the formation of precipitates on specific habit planes. The effectiveness of the thermal processing can be measured through aging curves,  which show how the alloy's hardness and mechanical strength gradually rise as a function of time at the annealing or aging temperature. The time coinciding with a maximum in hardness or strength is termed the ``peak-aged" condition; longer times lead to a decrease in hardness and strength. An examination of the resulting microstructure at the peak-aged condition reveals numerous precipitates. However, we claim that, in general, the size, shape, and distribution of these precipitates can be improved through the use of deformation in order to optimize hardness and strength. 
%This raises two questions: "What is the optimal precipitate structure?" and "How might we induce a structure closer to the optimal by different thermo-mechanical processing routes?" % However, there is a high probability that there are still some solute atoms left in the matrix.
\\

To assess the quality of the precipitate (-size) distribution and frequency, we begin by examining how solute atoms, solute clusters, and precipitates strengthen an alloy. Consider a case in which an alloyed sample is being mechanically deformed, resulting in a set of (mostly) basal dislocations traveling in a microstructure that contains obstacles, such as solute atoms, clusters, and precipitates. A moving dislocation encounters the solute atoms routinely within a maze of clusters and/or precipitates. The concentration of solute atoms, and the distribution and alignment of the clusters and precipitates determine how well these obstacles impeded the dislocation glide. Merely considering dislocations bowing around or shearing individual clusters and particles is insufficient. In addition, studies have shown that fine solute clusters and high densities of nanoscale precipitates can strengthen Mg alloys in ways that could not have been easily predicted from our knowledge of the effect of larger precipitates \citep{prameela2020deformation, ma2019dynamic}. Thus, we seek to address two related questions. First, regarding structure-property relationships: ``How do fine solute clusters or nanoscale precipitates contribute to high strength in Mg alloys, and, more generally, how do they influence the evolution of plastic microstructures?" Second, concerning processing-structure relationships: ``How can we most uniformly and reliably obtain fine solute clusters and nanoscale precipitates through processing?"
\\  

\begin{figure}[h]
   \centering
    \includegraphics[width=\linewidth]{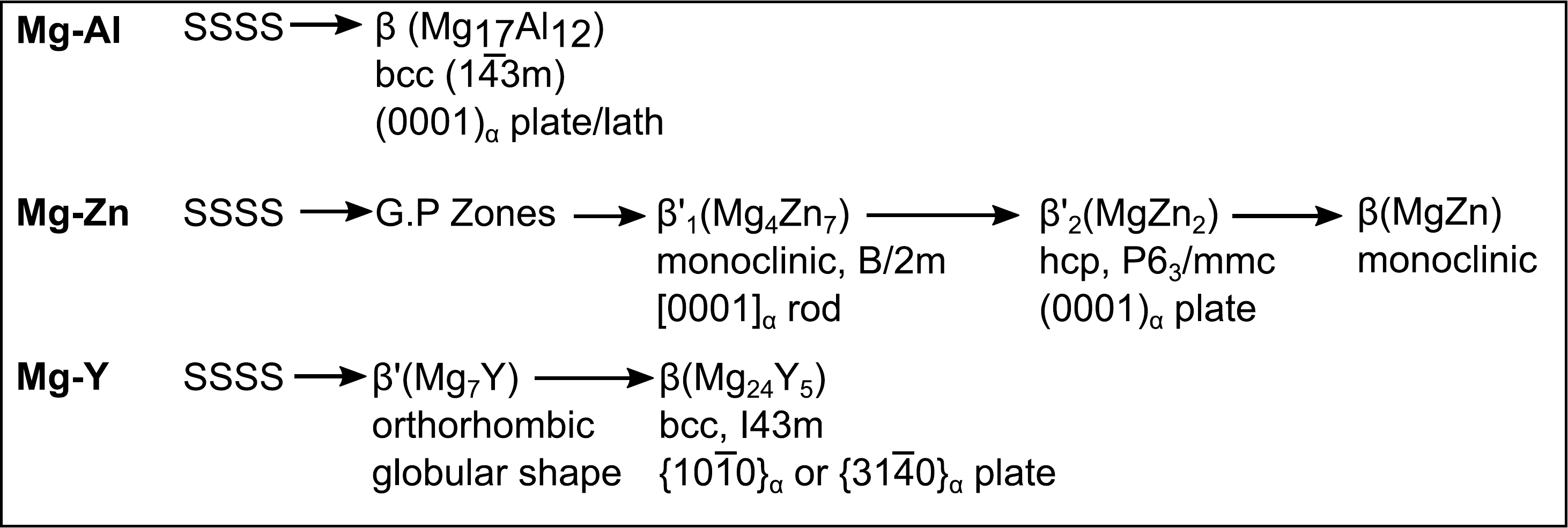}
    \caption{Precipitation sequence for Mg-Al, Mg-Zn, and Mg-Y systems, taken from \cite{22}}
    \label{fig:Process1}
\end{figure}

\begin{figure}[h]
    \centering
    \includegraphics[width=\linewidth]{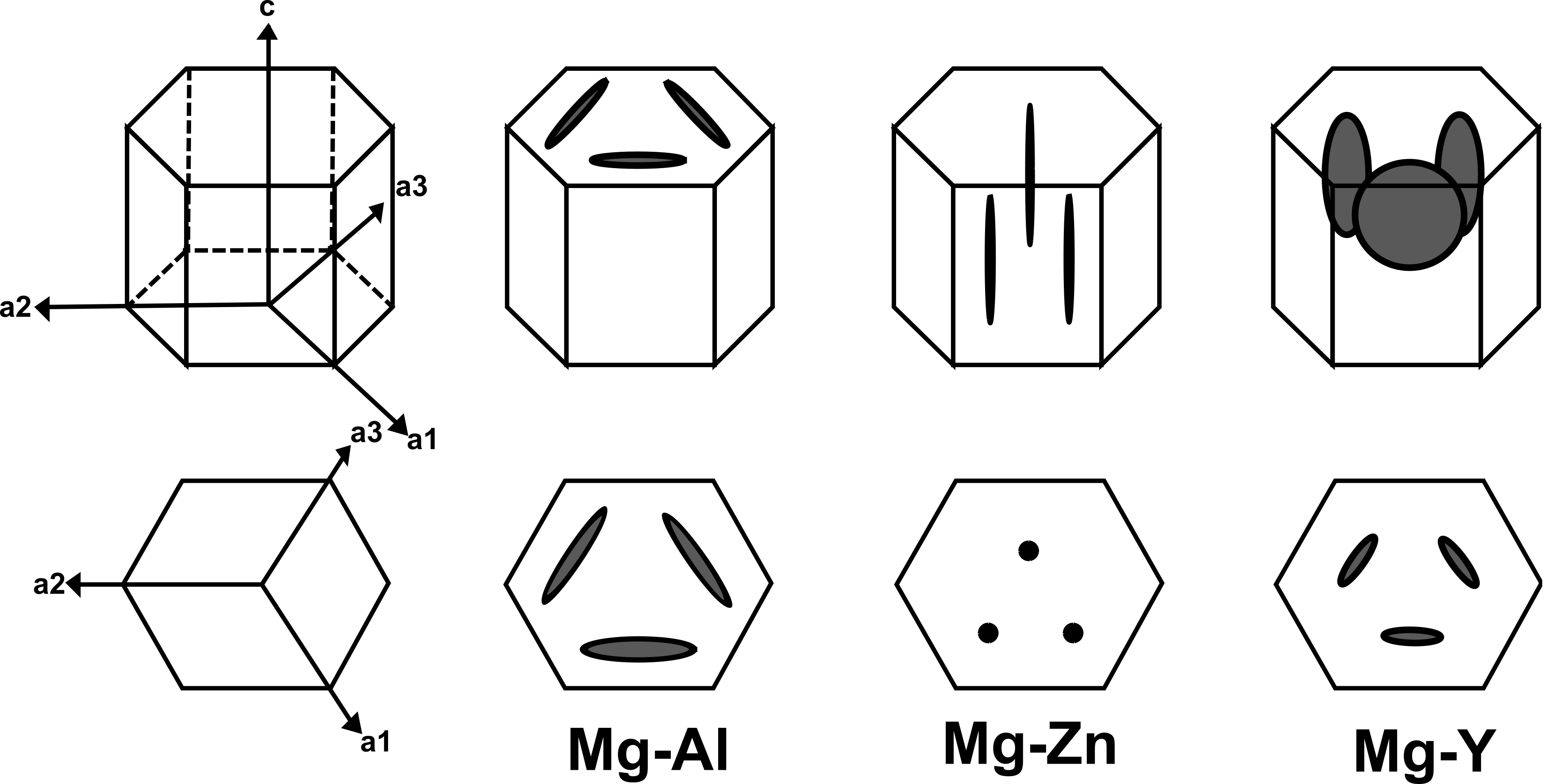}
    \caption{Perspective and top-down views of the precipitate morphology in Mg-Al, Mg-Zn and Mg-Y systems. }
    \label{fig:precipmorphology}
\end{figure}

\noindent In order to effectively address these questions, we focus on three binary Mg alloy systems as model materials: Magnesium-Aluminum(Mg-Al), Magnesium-Zinc(Mg-Zn), and Magnesium-Yttrium(Mg-Y). These binary alloys are far simpler than most commercial alloys, which contain multiple elements, but they provide clear guidance for many commercial alloy systems. For example, in the Mg-Al system, we consider two alloys: A6 and A9, which provide direct insight into the common commercial alloys AZ61 and AZ91, which each contain 1 wt.-\% Zn and other minor alloying elements. The three model binary systems also produce precipitates with distinct geometries, habit planes, and stoichiometries \citep{22}, as displayed in \cref{fig:Process1} and \cref{fig:precipmorphology}. We begin by reviewing the current understanding of how solute atoms, solute clusters, and precipitates control Mg's mechanical properties. We then explore the processing that is needed to generate the most effective array of precipitates and clusters. We end by considering an `integrated materials by design approach', which utilizes the understanding and methodologies obtained in these material systems for future processing design of other alloys.\\    

% \usepackage{multirow}
%\begin{table}[]
%\begin{tabular}{|l|l|l|l|l|}
%\caption{Melting point and maximum solubility of selected Magnesium alloy systems}
%\hline
%\multirow{2}{*}{\textbf{System}} & \multicolumn{2}{c|}{\textbf{Solubility}} & \multirow{2}{*}{\textbf{Precipitate Phase}} & \multirow{2}{*}{\textbf{Melting Point (°C)}} \\ \cline{2-3}
% & Wt.\% & At.\% &  &  \\ \hline
%Mg-Al & 12.7 & 11.6 & $Mg_{17}Al{12}$ & 402 \\ \hline
%Mg-Zn & 8.4 & 3.3 & MgZn & 347 \\ \hline
%Mg-Y & 12 & 3.6 & $Mg_{24}Y_{5}$ & 620 \\ \hline
%\end{tabular}
%\end{table}

Achieving “materials by design” has been a stated objective within the materials research community since at least the 1980’s  \citep{osti_6249241}. These aspirations have encompassed many aspects of traditional materials engineering including applying science-based approaches and, more recently, engaging in explicit multi-scale modeling. Enhanced computational capabilities that promise to establish clear causal connections have generated significant optimism regarding quantifying both how properties arise from atomic and mesoscale features and how microstructures arise from processing \citep{Olson995}. The Materials Genome Initiative \citep{national2011materials} and the National Research Council’s Integrated Computational Materials Engineering report \citep{national2008integrated} provided impetus to advance the computationally assisted design of materials, including metal alloys. In this regard, the ability to select and control the formation of complex  metallic microstructures via judicious choices of chemistry and carefully specified thermomechanical histories remains a grand challenge. \\

In their reviews of the Integrated Computational Materials Design (ICMD) process, \cite{xiong2015integrated, xiong2016cybermaterials} presented a systematic assessment of how this paradigm could be applied to alloy development, particularly non-ferrous alloys. A subsequent 2018 review of the U.S. Department of Energy funded PRISMS effort described some particular use cases wherein the ICMD process was applied to developing Mg-RE (rare-earth) alloys by closely integrating computational tools and state-of-the-art experiments  \citep{aagesen2018prisms}. These particular use cases included investigations guided by Density Field Theory (DFT)-informed Monte Carlo (MC) that detailed Guinier–Preston (GP) zone formation and the precipitation and growth of meta-stable phases from super-saturated solid solutions \citep{natarajan2016early, dewitt2017misfit, solomon2019stability}.  Additional use cases provide connections between precipitate structure and tensile strength in AZ91, investigated through phase field and dislocation dynamics (DD) modeling \citep{aagesen2018prediction}. More recently \cite{huang2021dislocation} also deployed MD models to inform strength models for the Mg-RE alloys . \\

In general, the work to date has sought to draw a straight line from chemistry to phase formation to microstructure to properties. However, \cite{xiong2015integrated} point out that one of the key developmental needs for the advancement of the ICMD paradigm is “further advances in linkage models of process–structure and structure–property relations.”  It is this inter-connectedness of processing and structure that we seek to address here; i.e. that mechanical deformation is not just controlled by microstructure but is often the process used for its sculpture.\\

In the ICMD vein, an international consortium within the Center for Materials under Extreme Dynamic Environments (CMEDE) has aimed to improve the performance of Mg alloys, among a multitude of other materials, under the paradigm ``Materials by design". A variety of strategies were developed to tackle imminent challenges and issues related to the use of Mg, be it for processing, hardening, the improvement of dynamic or static performance, the controlling of twins, precipitation, or approaches related to grain size effects. Experimental and modeling efforts were conducted and shared through the close collaborative network in CMEDE. The publications in this special volume reflect our collective results. \\

\section{Effect of solute atoms, solute clusters, and precipitates on plastic deformation and the evolution of microstructure}
\label{sec:EffectSolute}
\vspace*{1mm}

We start by considering the multiple ways that solute atoms, solute clusters, and nanoscale precipitates improve the mechanical properties of Mg alloys. They can impede the motion of dislocations and twins. They can also influence the generation and motion of dislocations and significantly influence overall texture evolution by suppressing recrystallization via grain boundary segregation \citep{robson2013effect}. Furthermore, precipitates can slow the motion of twins while also promoting the nucleation of additional twins.  Such interactions and mechanisms can significantly alter the yield surface of an alloy and its evolution during deformation \citep{wu2018mechanistic}. Thus, understanding the coupling of these microstructural features (solute atoms, solute clusters and precipitates) and plasticity carriers (dislocations, twins, grain boundaries) is critical to designing the microstructural features needed to optimize mechanical performance.  
Although we focus on solute clusters and precipitates, we include in our review the interaction between dislocations and randomly distributed solute atoms because (1) there are unique features of dislocation-solute interactions in Mg alloys, and (2) these interactions generate other defects like vacancies that play important roles in solute cluster and precipitate formation. \\

\subsection{Dislocation interactions}
\label{subsec:SoluteDislocation}
\vspace*{1mm}

\subsubsection{Interaction with solute atoms}
\label{subsec:soluteatomdislocation}
\vspace*{1mm}

The most commonly observed slip modes in Mg are $\langle a \rangle$ dislocation slip on basal and prismatic planes, and $\langle c+a \rangle$ dislocation slip on pyramidal planes \citep{zhang2012phenomenological}. These deformation modes react to randomly distributed solute atoms very differently.\\

\begin{figure}[h]
    \centering
    \includegraphics[width=\linewidth]{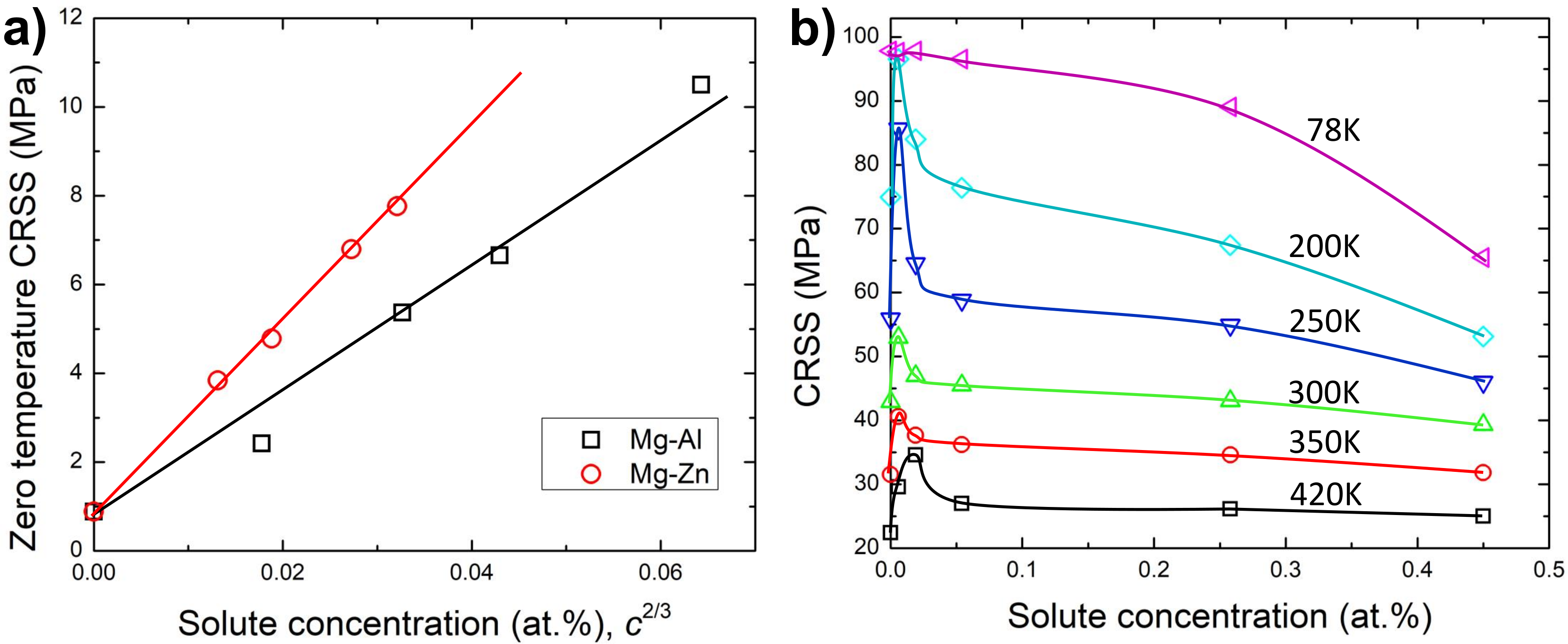}
    \caption{(a) Experimentally reported basal glide CRSS extrapolated to zero temperature for binary solid solutions of Mg-Al\citep{Akhtar1972}, Mg-Zn \citep{Akhtar1969} (lines are linear fit); (b) Experimentally reported temperature-dependent prismatic glide CRSS for binary solid solutions of Mg-Zn \citep{Akhtar1969} (curves are guide for the eye).}
    \label{fig:soluteeffects1}
\end{figure}

The most active slip mode, basal slip, displays typical solute strengthening behavior that is summarized in multiple reviews~\citep{Kocks1975,Argon2008}.  The increase of the CRSS scales with the solute concentration, $c$, as $c^k$ (\cref{fig:soluteeffects1}a).  The scaling exponent $k$=2/3 is most commonly observed for the hardening extrapolated to zero temperature \citep{Akhtar1972, RN1905,yi2016atomistic}, as predicted by the Labusch model and similarly to most FCC metals \citep{RN936, RN2109, RN1086, RN2094}.  According to the Labusch model, the zero temperature hardening can be expressed as \citep{RN962, RN947}

\begin{equation}
    \tau_{r0} = \left( {\frac{c^2vf_m^4}{b^3s^2 T_l}} \right)^{\frac{1}{3}},
\end{equation}

where $c$ is the solution concentration, $b$ is the Burgers vector, $s$ is the atomic area on the slip plane, $v$ is the solute-dislocation interaction range, $f_m$ is the interaction strength, and $T_l$ is the dislocation line tension.  Other models predict values of the scaling parameter $k$ to be between $1/2$ and 1, depending on solute concentration, dislocation tension, and the strength of solute-dislocations \citep{Akhtar1972, RN995, RN962, RN947, RN2094}.
To gain some insight into alloy choices, \cite{RN2094} demonstrated that, in the dilute limit and counting only the elasticity contribution of the solute-dislocation interaction, zero-temperature solute strengthening is proportional to the misfit volume of the solute atoms, i.e.,
\begin{equation}
    \tau_{r0} \propto c^\frac{2}{3} {\Delta V}^{\frac{4}{3}}.
\end{equation}
According to \cite{RN1018}, the temperature dependence of the CRSS can be divided into four temperature regimes.  At low temperatures (or high strain rates), dislocations glide by inertia. At high temperatures (or low strain rates), dislocations are constantly surrounded by solute clouds due to the solute atoms' high mobility.  In the medium-temperature range, dislocation slip is mostly controlled by the rate of overcoming solute barriers by thermal activation.  The temperature dependence of the basal slip in the medium temperature regime was shown by \cite{RN1905} to be expressed by
\begin{equation}
  \tau_{c}(T)=\begin{cases}
    \tau_{r0}\left[ 1- \left( \frac{k_BT}{\Delta E_b}\text{ln}\frac{\dot \epsilon_0}{\dot \epsilon} \right)^{\frac{2}{3}}\right], & \text{if $\frac{\tau_{c}(T)}{\tau_{c0}} > 0.5$}.\\
     \tau_{r0}\text{exp}\left( -\frac{1}{C_l}\frac{k_BT}{\Delta E_b}\text{ln}\frac{\dot \epsilon_0}{\dot \epsilon} \right), & \text{if $0.2 < \frac{\tau_{c}(T)}{\tau_{c0}} < 0.5$},
  \end{cases}
\end{equation}
where $\Delta E_b$ is the solute barrier height, parameter $C_l$ is about $0.55$, and $\dot \epsilon_0$ is a fitting parameter.  Finite temperature leads to some change in the scaling parameter $k$ that is observed in experiments and molecular dynamics (MD) simulations \citep{Akhtar1972, yi2016atomistic}.\\

While basal slip remains in-plane, out-of-plane dislocation motion is common for other slip modes in Mg solid solutions.  For prismatic slip, the screw $\langle a \rangle$ dislocation glides by cross-slip between the prismatic plane and the basal plane in a locking-unlocking manner at low temperature and a wavy manner at high temperature \citep{RN1106, RN1107, RN1103}.  Solute softening was observed at low temperatures (\cref{fig:soluteeffects1}b) due to the solute atoms facilitating kink-pair formation \citep{RN1104, RN948, RN1073, yi2017solute}. This behavior has also been reported in many body-centered cubic (BCC) metals \citep{RN2104, RN2101, RN1071}. \\

\begin{figure}[h]
    \centering
    
    \includegraphics[width=\linewidth]{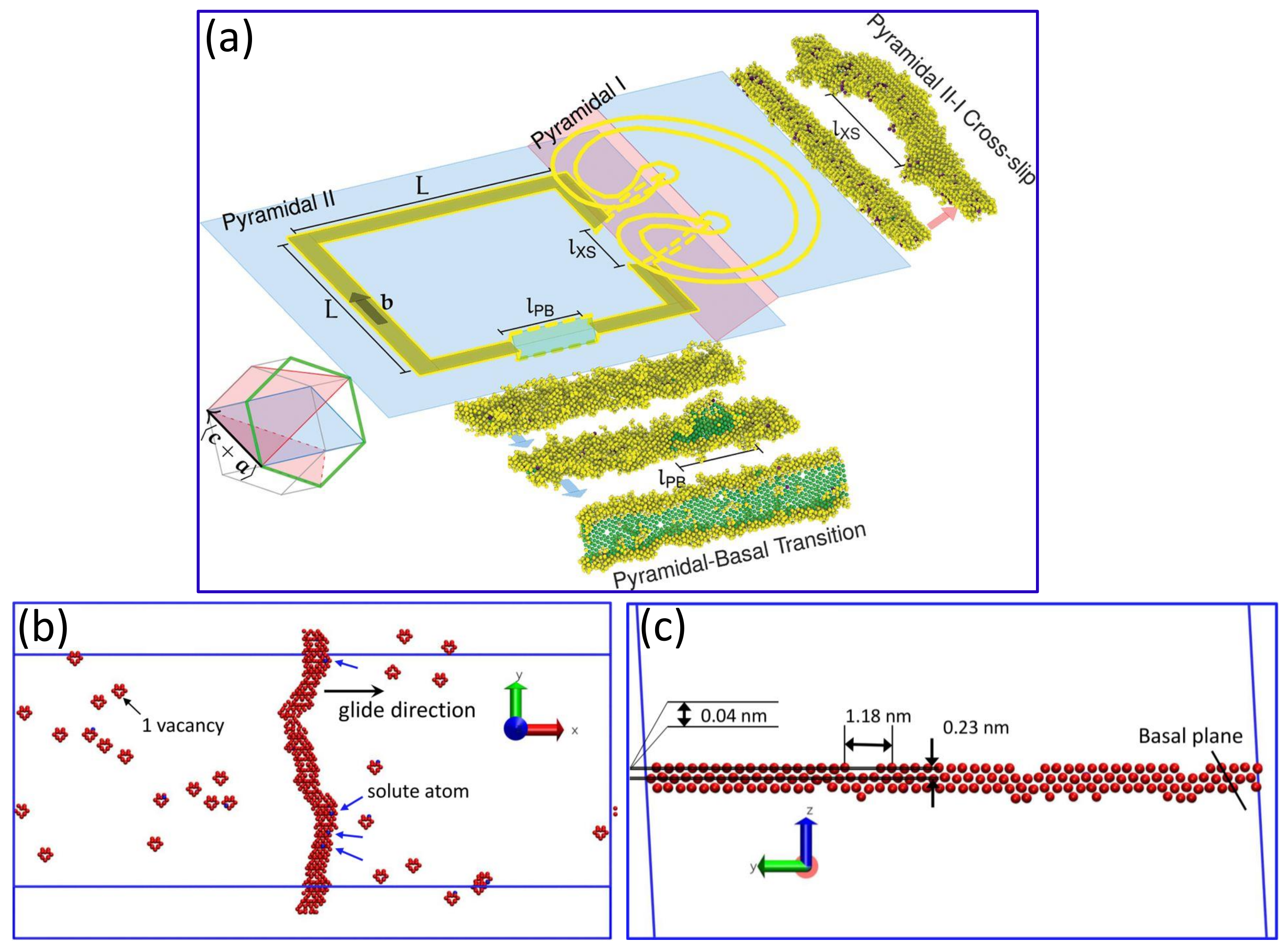}
    \caption{(a) MD simulation of double cross-slip of $\langle c+a \rangle$ dislocation to form Frank-Reed sources \citep{RN2052}; (permission request being processed)
    (b)(c) Molecular dynamics simulation of climb for an $\langle c \rangle$ edge dislocation on the pyramidal I plane.\citep{YIclimb2021}
   Reprinted from Scripta Materialia, Vol 194, Peng Yi, Solute softening and vacancy generation by diffusion-less dislocation climb in Mg alloys, Copyright 2021, with permission from Elsevier.   } 
    %Reprinted from Scripta Materialia, Vol 194, Peng Yi, Solute softening and vacancy generation by diffusion-less dislocation climb in Magnesium alloys, Pages No., Copyright 2021, with permission from Elsevier.
    \label{fig:soluteeffects2}
\end{figure}

Cross-slip is also essential in $\langle c+a \rangle$ dislocation glide, which is arguably the primary deformation mode accommodating deformation along the $\langle c \rangle$-axis of the HCP structure, but which is also the most difficult to activate. Simulations only recently suggested that $\langle c+a \rangle$ dislocations double cross-slip between pyramidal I and pyramidal II planes to form Frank-Reed sources that lead to multiplication, as shown in \cref{fig:soluteeffects2}a \citep{RN2050, RN2052}.  Therefore, the solute softening mechanism can potentially improve the mobility of $\langle c+a \rangle$ slip and therefore reduce the slip mode anisotropy \citep{RN1246, RN2096}.\\

Another type of out-of-plane dislocation motion, climb, is traditionally associated with high-temperature creep.  Direct experimental studies of climb in Mg alloys are scarce \citep{Caillard2007}. A recent MD simulation study revealed active climb of non-basal $\langle a \rangle$ dislocations at room temperature \citep{YIclimb2021}  (\cref{fig:soluteeffects2}b). The observed solute-facilitated climb leads to softening and reduced CRSS anisotropy, and may improve the ductility of Mg materials.  In addition, climb of $\langle c \rangle$ dislocations could offer an alternative to $\langle c+a \rangle$ dislocations in accommodating the $\langle c \rangle$ axis deformation \citep{RN2057}.  Both out-of-plane motions, cross-slip and climb, generate an enormous number of vacancies \citep{yi2017solute,YIclimb2021}, in addition to the usual vacancy generation by jogs formed when dislocations cross each other.  Vacancies impact not only the mechanical properties, but also the precipitation process as discussed in \cref{vacancyhetnucleation}.\\

Understanding the effects of solute atoms on different slip modes enables the design of alloys with more isotropic deformation modes, which may improve the Mg alloys' ductility.  For example, prismatic slip softening often reaches a maximum at a specific solute concentration, and, in Mg-Zn alloys, the ductility exhibits a maximum at the same concentration \citep{RN2110}.\\

\begin{figure*}
    \centering
    \includegraphics[width=0.8\linewidth]{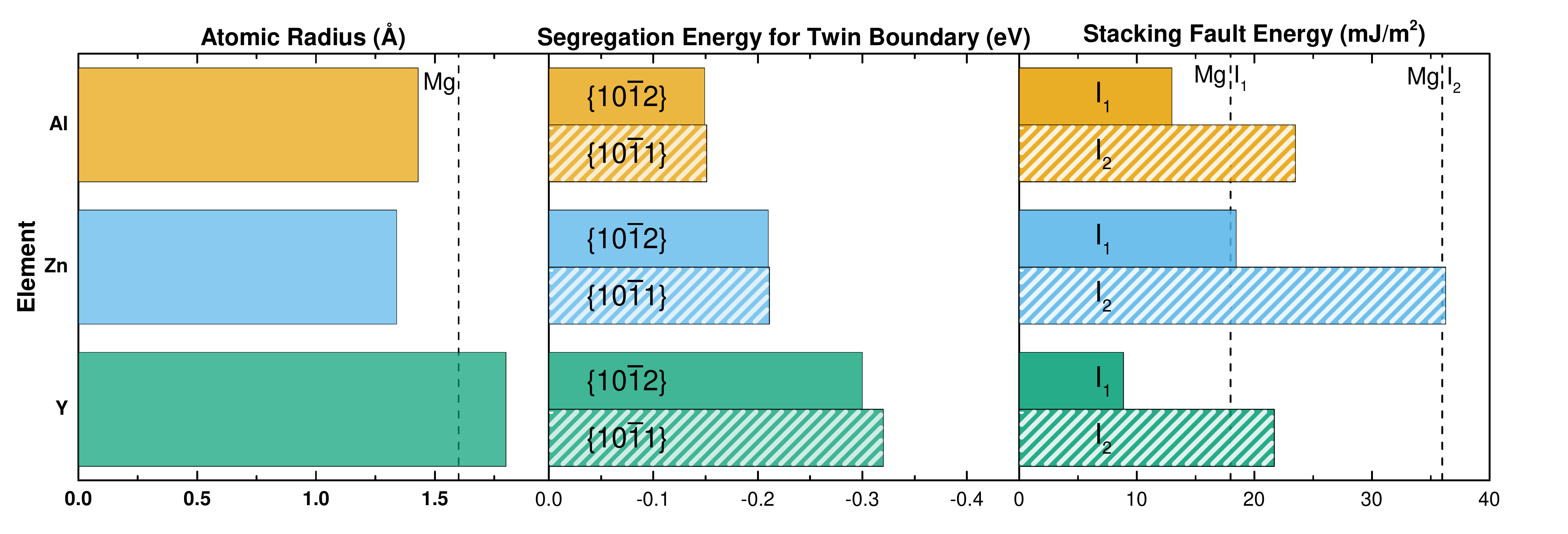}
    \caption{Segregation energy and stacking fault energy of Al, Zn, and Y. Segregation energy data from \citep{SEref1,SEref2,SEref3,SEref4,SEref5}, stacking fault energy data from \citep{SFEref1,SFEref2,SFEref3,SFEref4,SFEref5}. Figure adapted from \cite{Nie2020SESFE}.}
    \label{fig:SFEandSE}
\end{figure*}

While the addition of solute atoms can alter CRSS values, the presence of dislocations or stacking faults can, in turn, alter solute distributions in important ways for Mg-Al, Mg-Zn and Mg-Y alloys (\cref{fig:SFEandSE}). A static or slow-moving dislocation can draw a solute cluster, or solute cloud, since this redistribution of solute reduces the strain energy around the dislocation.  When the shear stress is low, a solute cluster may pin its dislocation.  With increasing shear stress, the dislocation will break free until it again remains sufficiently stationary to attract a new solute cluster.  Solute clusters are constantly forming and dissolving around static or slow-moving dislocations. The interplay between solute cluster formation and dislocation motion, known as the solute drag effect, contributes importantly to the dynamic strain aging process. This problem was discussed recently by \citep{12,13}. In \cref{dislocationhetnucleation} we discuss how dynamic processing methods may promote  nanoscale precipitation since these methods induce and multiply dislocations, as well as other types of defects, that attract solute atoms. The resulting solute segregation may serve as dynamically induced heterogeneous nucleation sites that assist phase transformation.\\

\subsubsection{Interaction with solute clusters}
\label{dislocation_soluteCluster_interaction}
\vspace*{1mm}

Compared to the randomly distributed solute atoms in a solid solution, solute clusters are localized regions with much higher concentrations than the average in the solid solution.  Typical solute clusters contain $10^1-10^2$ atoms.  They often form in a super-saturated solid solution during thermal processing, as precursors to the second phase precipitates; a well-known example of this is the GP zones.  Solute clusters could also form due to interaction with defects, such as the solute clouds segregated around dislocations mentioned in \cref{subsec:soluteatomdislocation}.\\

Solute cluster strengthening was recognized in Al alloys as early as in the 1960s, notably associated with the two-stage hardening curves ~\citep{Baba1969,FAN2018152}; however, systematic study of solute cluster strengthening started only in the past two decades, with the development of high resolution imaging and characterization methods like APT and HAADF-STEM ~\citep{Marceau2016,Dumitraschkewitz2018}.  The studies are most extensive in Al alloys, but have been extended into Mg alloys \citep{Marceau2016}.\\

Since solute clusters have much higher solute concentration than the average in the solid solution while still maintaining the matrix crystal structure, they provide a different strengthening mechanism from solid solution strengthening or precipitate strengthening, the latter of which is discussed in \cref{dislocation_precipitate_interaction}.
As compared to individual solute atoms, solute clusters are much stronger obstacles. Therefore, Friedel statistics are more applicable than the Labusch statistics used for solid solutions~\citep{RN995}. In contrast to precipitates, solute clusters are always penetrable (shearable), while precipitates can be penetrable or impenetrable.  Under many circumstances, precipitates are considered impenetrable and follow a specific strengthening mechanism called Orowan mechanism \citep{15}. Historically, the systematic study of precipitates precedes that of solute clusters because precipitates are much easier to observe experimentally.  We present solute cluster strengthening before the precipitate strengthening, following the progression of the length scale.\\

\cite{DESCHAMPS20112918} first adopted Friedel statistics to model the strengthening effect of solute clusters in Al alloys based on their experimental characterization using \textit{in situ} Small-angle X-ray Diffraction(SAXS) and Nuclear Magnetic Resonance(NMR) techniques. Friedel statistics~\citep{1964Friedel} express the strengthening effect as~\citep{RN962, RN947}

\begin{equation}
    \tau_{r0} = \frac{f_m^{3/2}\sqrt{c}}{b\sqrt{2sT_l}}, 
    \label{Eq:FriedelStatistics}
\end{equation}

where $f_m$ is the average obstacle strength of the solute clusters, and $c$ is the concentration of clusters in the system.  With a fixed solute concentration, solute clustering increases $f_m$ but reduces $c$.  This interplay of these two factors leads to the aforementioned two-stage hardening curve observed in their experiments.\\

\cite{ZHAO201443} considered clusters' elastic modulus effect and proposed a size misfit strengthening model, where the size misfit of clusters is assumed to be the sum of elementary atomic misfits. The predictions of this model were in good agreement with experiments for small clusters ($<$ 1 nm) in Al alloys. \cite{DEVAUCORBEIL2013b} studied the effects of non-randomly distributed solute atoms on the CRSS of dislocation glide.  The effects show three regimes dependent on the obstacle strength of individual solute atoms.  If the obstacle strength is very weak, then clustering has no effect.  In the intermediate regime, the clusters have the same effect as impenetrable precipitates (see \cref{dislocation_precipitate_interaction}).  In the limit of infinitely strong obstacles, clustering lowers the overall strength.  Several of the same authors \citep{MARCEAU20137285} further developed the model to directly incorporate the cluster distribution determined by APT into a randomly arranged areal glide plane model for strengthening prediction.  The strength of each cluster is assumed to be proportional to its radius on the glide plane, and is summed using a newly proposed relation for the superposition of strengthening components~\citep{DEVAUCORBEIL2013a}.\\

\begin{figure}[h]
    \centering
    \includegraphics[width=0.9\linewidth]{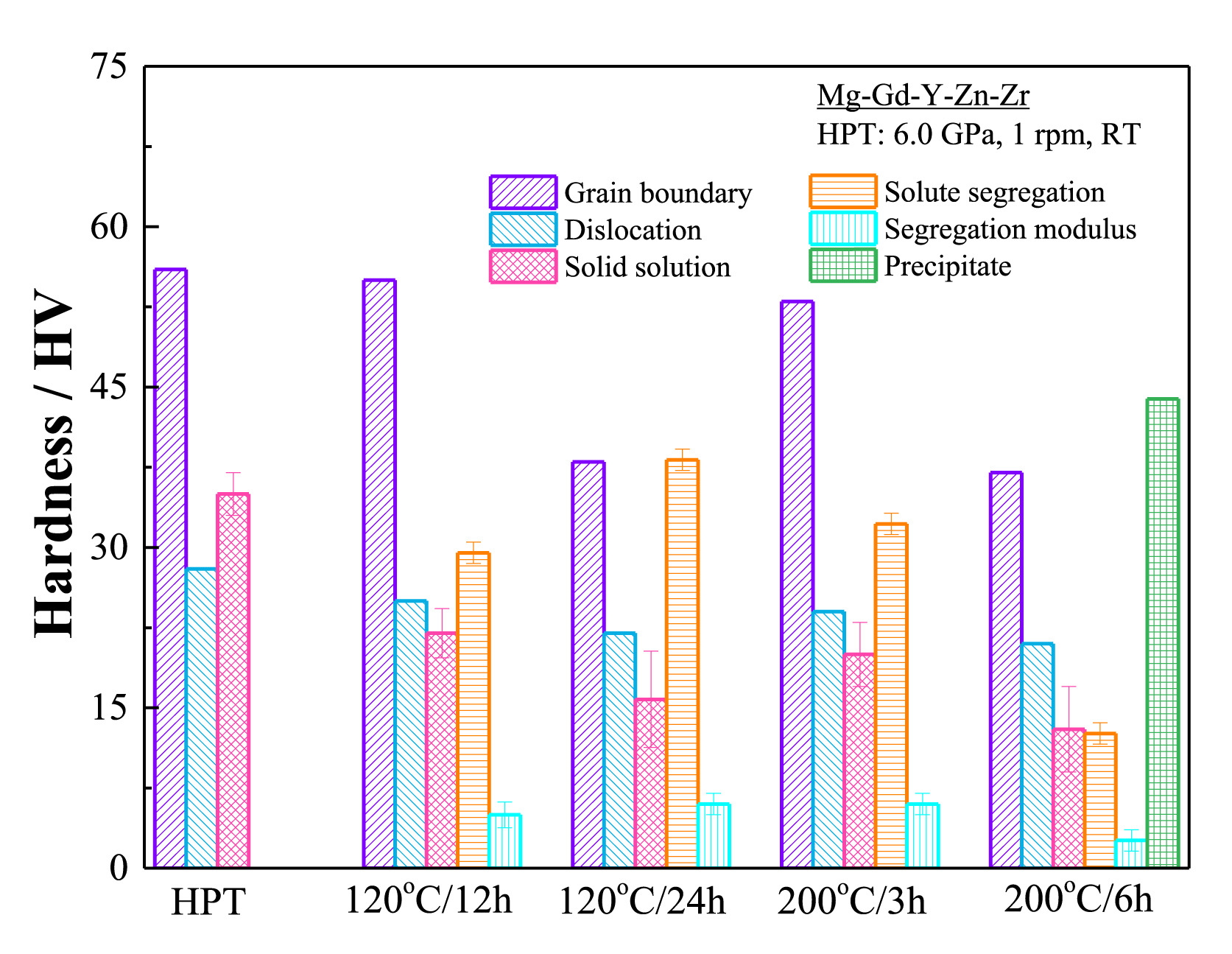}
    \caption{Contribution of each mechanism to hardness for HPT-processed Mg-Gd-Y-Zn-Zr alloy under different ageing conditions;% (b) Comparison between calculated hardness and measured hardness.  
    Figure taken from  \cite{SUN2018260}.Reprinted from Acta Materialia , Vol 151, W.T. Sun,X.G. Qiao,M.Y. Zheng,C. Xu,S. Kamado,X.J. Zhao,H.W. Chen,N. Gao,M.J. Starink, Altered ageing behaviour of a nanostructured Mg-8.2Gd-3.8Y-1.0Zn-0.4Zr alloy processed by high pressure torsion, Pages No. 260-270, Copyright 2018, with permission from Elsevier.}
    \label{fig:soluteClusterHardening}
\end{figure}

Starink et al. \citep{STARINK20092376,STARINK20124194} proposed a thermodynamic model based on a single interaction energy of dissimilar nearest neighbors, and includes treatment of short-range order hardening, elastic modulus hardening and chemical hardening. \cite{SUN2018260} have applied this model to predict the hardness of Mg-Gd-Y-Zn-Zr alloys processed by high pressure torsion (HPT) and concluded that the main strengthening mechanism in the peak-aged samples was solute cluster strengthening instead of precipitates strengthening (\cref{fig:soluteClusterHardening}).  \cite{MO2019185} also examined the Mg-Gd-Ca alloys and observed a significant strengthening contribution from the Gd-Ca solute clusters.  Similar to precipitation hardening, a high number density of solute clusters with small inter-cluster spacing can provide higher strengthening effects.\\

\subsubsection{Interaction with precipitates}
\label{dislocation_precipitate_interaction}
\vspace*{1mm}
Precipitates also interact strongly with dislocations in alloys and can impede the dislocations' motion~\citep{Ardell1985,Argon2008}.  Unlike the precipitates in cubic alloys, precipitates in Mg alloys tend to elongate along different crystallographic directions, as shown in \cref{fig:precipmorphology}.  Understanding how precipitates' habit planes, densities, and relative spacings impact their ability to impede slip is essential for Mg alloy design.\\

Two limiting cases for precipitate strengthening have been identified depending on whether dislocations can penetrate the precipitates.  For impenetrable precipitates, the dislocation bypasses the obstacles and leaves loops around the particle.  This is the so-called Orowan mechanism. Penetrable precipitates, on the other hand, act like solute clusters, and therefore Friedel statistics apply, as introduced in \cref{dislocation_soluteCluster_interaction}.\\

The CRSS, as originally proposed by Orowan~\citep{Orowan1948symposium}, can be expressed as $\tau_{Orowan}=G_mb/L$, where $G_m$ is the shear modulus of the matrix, $b$ is the Burgers vector and $L$ is the spacing between precipitates.  There are several modifications throughout the years that differ slightly in forms or numerical factors but all capture the same physics~\citep{Queyreau2010}.  Based on \cite{Hirsch1969}, the CRSS can be expressed as

\begin{equation}
%\label{eq: 5}
    \tau _{Orowan} = \frac{G_m b}{2\pi (d_s - 2r_p)\sqrt{1-\nu}} \text{ln}\frac{2 r_p}{r_0},
    \label{OrowanEquation}
\end{equation}
where $r_0$ is the core radius of the dislocation \citep{Kocks1975}, $r_p$ is the average radius of the precipitates on the slip plane,  $\nu$ is Poisson's ratio, $d_s$ is the spacing between the precipitates on the glide plane, and  $d_s=n_s^{-1/2}$ where $n_s$ denotes the number of precipitates per unit area.\\

\begin{figure}[h]
    \centering
    \includegraphics[width=\linewidth]{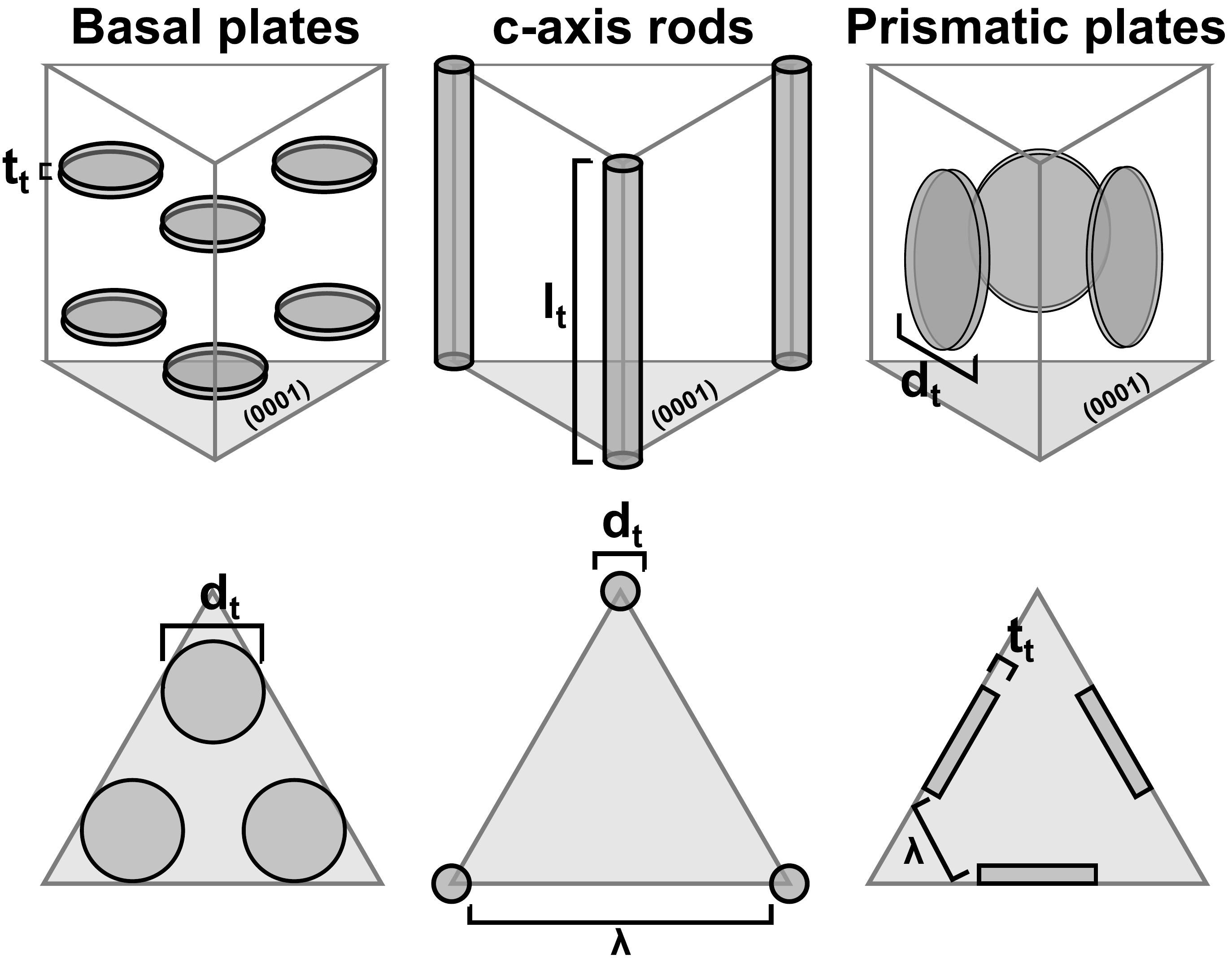}
    \caption{Schematic of geometric terms used in Orowan Equation adaptations for basal plates, c-axis rods, and prismatic plates.}
    \label{fig:OrowanTerms}
\end{figure}

\cite{17} and others have investigated the effect of precipitate hardening of Mg  under conditions dominated by the Orowan mechanism. They calculated the effective planar inter-precipitate spacing $\lambda _e=d_s-2r_p$ of different precipitate morphologies in the HCP system (\cref{fig:OrowanTerms}) and estimated the change in $\tau _{Orowan}$ using %equation [Orowan equation]
\cref{OrowanEquation} above.  \\

Let $f$ denote the volume fraction of the precipitates, and assume that $r_0=b$. The Orowan CRSS for basal plates, prismatic plates, and $c$-axis rods, are, respectively, given by
\begin{equation}
    \tau _{Orowan,b}=\frac{G_m b}{2\pi\left(\frac{0.953}{\sqrt{f}}-1\right) d_t \sqrt{1-\nu}}\text{ln}\frac{d_t}{b},
    \label{Eq:basalOrowan}
\end{equation}
\begin{equation}
    \tau_{Orowan,p}=\frac{G_m b \text{ ln}(\frac{0.886\sqrt{d_t t_t}}{b})   [ \sqrt{1-\nu}]^{-1}}{2\pi (0.825\sqrt{\frac{d_t t_t}{f}}-0.393d_t-0.886t_t)},
    \label{Eq:prismaticOrowan}
\end{equation}
and
\begin{equation}
    \tau_{Orowan,c}=\frac{G_m b}{2\pi\left(\frac{0.953}{\sqrt{f}}-1\right)d_t \sqrt{1-\nu}}\text{ln}\frac{d_t}{b},
    \label{Eq:c-axisOrowan}
\end{equation}

where the meanings of parameters $d_t$ and $t_t$ are illustrated in \cref{fig:OrowanTerms}. It follows from~\cref{Eq:basalOrowan,Eq:prismaticOrowan,Eq:c-axisOrowan} that $\tau_{r}$ depends strongly on $f$ and precipitate diameter ($d_t$), which in turn depend on the effective inter-particle spacing $\lambda_e$.\\  

\begin{figure}[h]
    \centering
    \includegraphics[width=\linewidth]{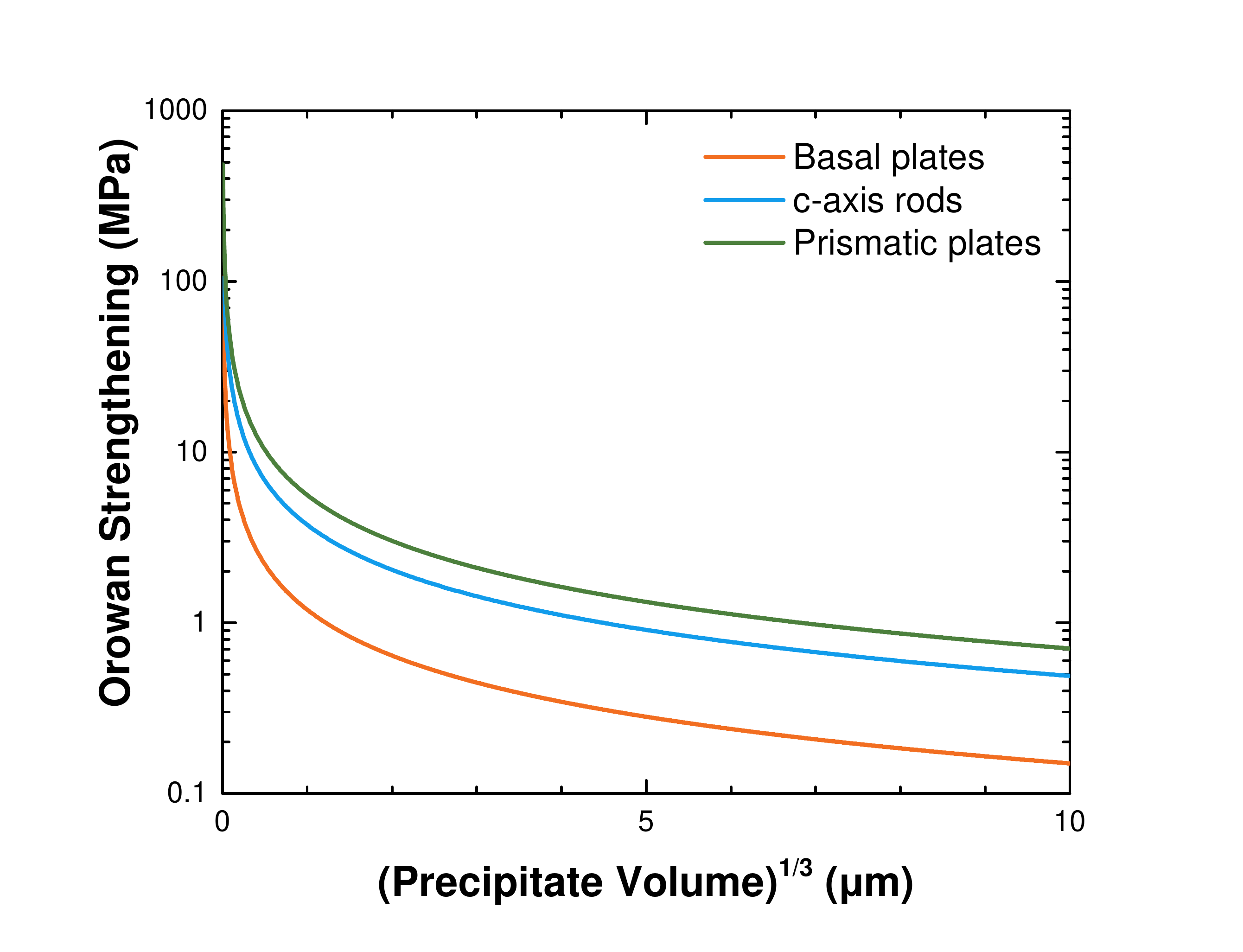}
    \caption{Calculated Orowan stress required to bypass basal plates, prismatic plates, and $c$-axis rods. Particle volume fraction is 5 \%, aspect ratio of plates 0.1, and aspect ratio of rods 10. Calculations as per \cite{17}.}
    \label{fig:OrowanStresses}
\end{figure}

Based on \cref{Eq:basalOrowan,Eq:prismaticOrowan,Eq:c-axisOrowan}, \cref{fig:OrowanStresses} plots the Orowan stress as a function of the effective linear dimension of the precipitates to compare different shapes and habit planes at the same total volume fraction and number density of the precipitates. The comparison shows that the prismatic plates provide maximum strengthening for Mg alloys. However, prismatic plates are generally formed by adding Rare-earth (RE) elements (such as Y, Nd, Gd) to the Mg matrix. These RE elements are expensive and not widely available. The scarcity of RE alloying elements thus limits their extensive practical use. Hence, there is a need to focus beyond RE elements in strengthening Mg alloys. A suitable alternative focuses on increasing the number density of basal or $c$-axis rods while decreasing their spacing to increase strength. This focus directs us towards a high density of nanoscale precipitates in \cref{fig:OrowanStresses}, which brings an additional benefit. Recent studies have shown that nanoscale precipitates not only increase strength, but also enhance ductility \citep{peng2020nanoscale}.\\

\cref{fig:OrowanStresses} suggests that the Orowan stress continues to increase with decreasing precipitate size.  On the other hand, the stress required to shear the precipitates decreases with decreasing precipitate size.  At some critical precipitate size, the shearing CRSS is lower than the Orowan stress and the strengthening mechanism transitions from the Orowan mechanism to Friedel statistics.  In some Mg alloys, the penetrable precipitate model offers better agreement with experiments than the impenetrable precipitate model~\citep{CEPEDAJIMENEZ2019164,SOLOMON201867}.\\

%When the dislocation is able to cut through the precipitates ($\phi_c>0$),  \cite{1964Friedel} proposed that the critical resolved shear stress can be expressed as
%\begin{equation}
%    \tau _r= 2 \frac{\alpha G_m b^2}{bd_s} \cos^{3/2}\phi _c.
%    \label{PenetrablePrecEquation}
%\end{equation}

The value of this critical precipitate size depends on a number of factors, like matrix-precipitate interface coherency and applied stress.  But the critical value generally falls into the range of 50-100$b$, or 15-30 nm~\citep{FAN2018152}. More recent precipitate hardening models were proposed to consider the cases where both impenetrable and penetrable mechanisms are in play~\citep{BHATTACHARYYA201855,Esteban_Manzanares_2019, FAN2018152}.  Whether or not the precipitates are penetrable, it is worth noting that decreasing the precipitate spacing, i.e., increasing the precipitate number density, leads to improved precipitate strengthening for a given precipitate volume fraction (\cref{Eq:FriedelStatistics,OrowanEquation}).\\

\subsection{Twin interactions}
\label{subsec:SoluteTwin}
\vspace*{1mm}

Twinning is common in Mg because dislocations that can accommodate $c$-axis deformation, including $\langle c+a \rangle$ and $\langle c \rangle$ dislocations, are difficult to activate \citep{RN982, RN1131, zhang2012phenomenological}.  The most prevalent twin modes are the tension twin and compression twin, with the former being dominant \citep{RN2114, RN2113}.  Tension twinning is the second easiest deformation mode in Mg after basal slip (as shown in \cref{fig:DeformationMechanisms}) , and is the most studied mode of deformation.\\

\subsubsection{Interaction with solute atoms and solute clusters}
\begin{figure}[h]
    \centering
    \includegraphics[width=0.55\linewidth]{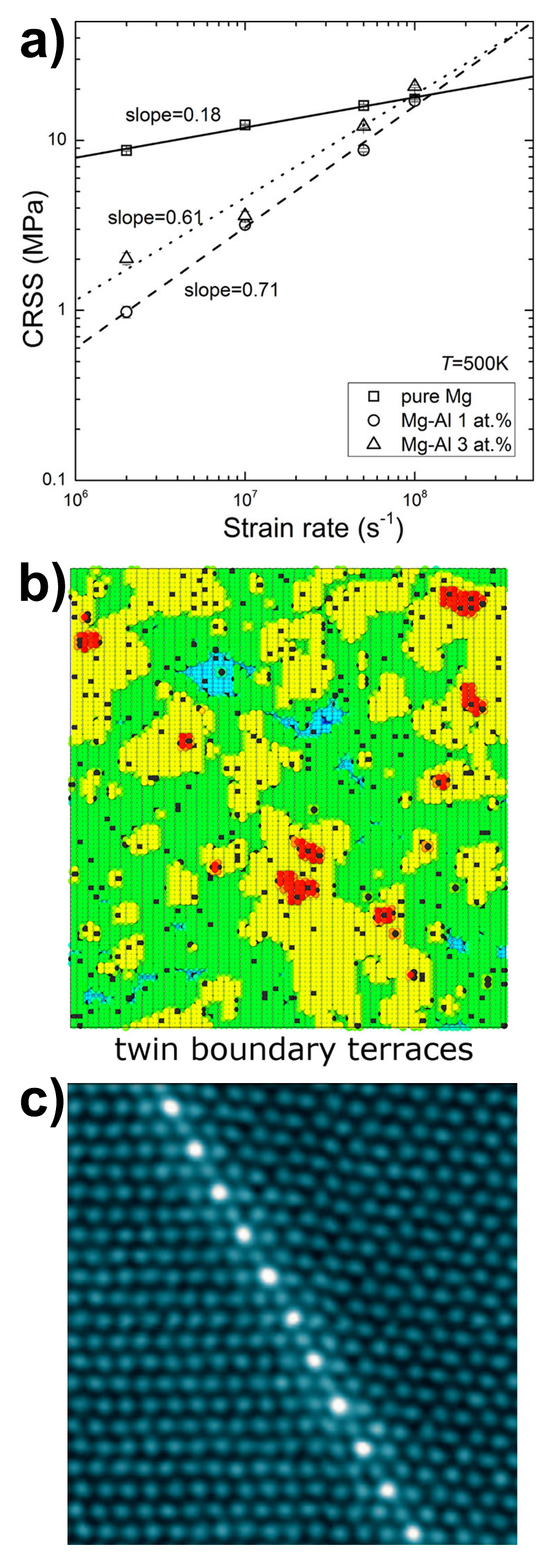}
    \caption{(a) Strain rate sensitivity of the CRSS of tension twin thickening in a Mg/Al alloy. (b) A snapshot of solute atoms on the tension twin boundary, which consists of several planes in Mg/Al alloys. Figures (a) and (b) taken from \citep{Yi2019}, with permission from Elsevier.
    %    Reprinted from Scripta Materialia, Vol 162, Pengi Yi, Michael Falk, Thermally activated twin thickening and solute softening in Magnesium alloys - a molecular simulation study, Pages No., Copyright 2019, with permission from Elsevier.
        (c) Solute segregation along twin boundaries. The figure is taken from \citep{18}. } 
    \label{fig:Yi2019Nie2013}
\end{figure}

\vspace*{1mm}
Although twins are three-dimensional (3D) defects, most studies, with few exceptions \citep{RN2126, RN2127}, have focused on the two-dimensional (2D) plane believed to dominate the twin nucleation and growth.  While studies of twin nucleation are relatively rare ~\citep{RN2120, RN2122, RN1236}, most work on twin mobility has been devoted to twin growth \citep{RN1237, RN2119, RN2097, RN1133, RN890, RN917, RN2124, RN2125}. The growth along the normal to the twin plane, called thickening, proceeds through nucleation and expansion of twin dislocation (TD, also called twin disconnection) loops, where solute atoms facilitate the nucleation of TD loops and resist their expansion \citep{RN916, RN917, Yi2019}.  The interplay between the nucleation and expansion processes results in temperature- and strain rate-dependent solute effects \citep{Yi2019} (see \cref{fig:Yi2019Nie2013}(a)(b)).  Growth parallel to the twin plane, called propagation, is also accommodated by TDs, which form basal-prismatic (BP) interfaces, prismatic-basal (PB) interfaces, and conjugate twin boundaries \citep{RN2097, RN2124}.  During twin propagation, solute atoms are expected to form barriers that resist the propagation of the BP and PB interfaces.   It is worth mentioning that recent high-resolution TEM experiments reported solute segregation near strained twin boundaries, which provides a pinning effect and resembles the solute cloud near dislocations (see \cref{fig:Yi2019Nie2013}(c)). Many similarities are expected between solute-twin interactions and solute-dislocation interactions.  Finally, as studies on solute clusters in Mg alloys are still at a relatively early stage, we anticipate forthcoming information regarding the interaction between solute clusters and twins in Mg alloys soon.\\

\subsubsection{Interaction with precipitates}
\vspace*{1mm}

\begin{figure}[h]
    \centering
    \includegraphics[width=\linewidth]{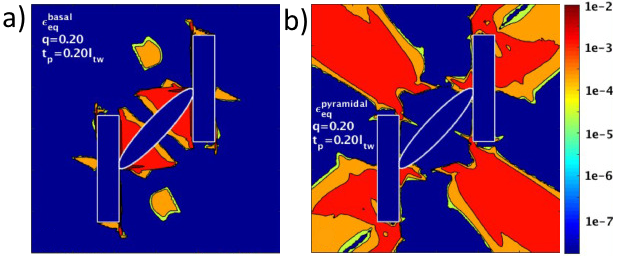}
    
    \caption{Map of equivalent  plastic strain produced by (a) basal and (b) pyramidal slip around a twin with aspect ratio 0.2. Figure from \citep{Siska2021}. Reprinted from Acta Materialia, Vol 202, Filip Siska, Ludek Stratil, Jan Cizek, Tingting Guo, Matthew Barnett, Numerical analysis of twin-precipitate interactions in Magnesium alloys, 2021, with permission from Elsevier. }

    %Reprinted from Acta Materialia, Vol 202, Filip Siska, Ludek Stratil, Jan Cizek, Tingting Guo, Matthew Barnett, Numerical analysis of twin-precipitate interactions in Magnesium alloys, Pages No., 2021, with permission from Elsevier.
    \label{fig:TwinsPrecip}
\end{figure}

In general, precipitate-twin interactions are less understood than precipitate-dislocation interactions \citep{RN2117, RN2116, RN2118, robson2013effect}. When twins approach precipitates that can't be sheared, the twins are blocked and must bow around the particle (see \cref{fig:TwinsPrecip}). In this process, the twins engulf and rotate the precipitate. In other cases, precipitates can be sheared by twins ~\citep{clark1965transmission, ZHANG2020203}, although these cases are thought to be less common, and more study is needed. Some recent studies show interesting results that may lead to a better understanding of the precipitate-twin interaction, such as that written by \cite{liu2018integrated}. In their single- and poly-crystal simulations, the authors investigated the effect of basal plate precipitates. Their investigations regarding the interaction of a pair of basal plate precipitates with twins showed that these precipitates have a strong hardening effect by impeding the twin domain propagation. The analysis of larger and smaller precipitates at lower and higher number densities, respectively, showed that precipitate volume fraction, number density, and orientation play important roles in precipitate strengthening. 
It was found that plates oriented parallel to the twin boundary show the strongest effect. Precipitates oriented in basal or prismatic habit planes lead to a lower hardening effect. It was further found that a large number of small precipitates has a larger impact on the hardening behavior than a small number of large precipitates.\\

\begin{figure}[b]
    \centering
    \includegraphics[width=\linewidth]{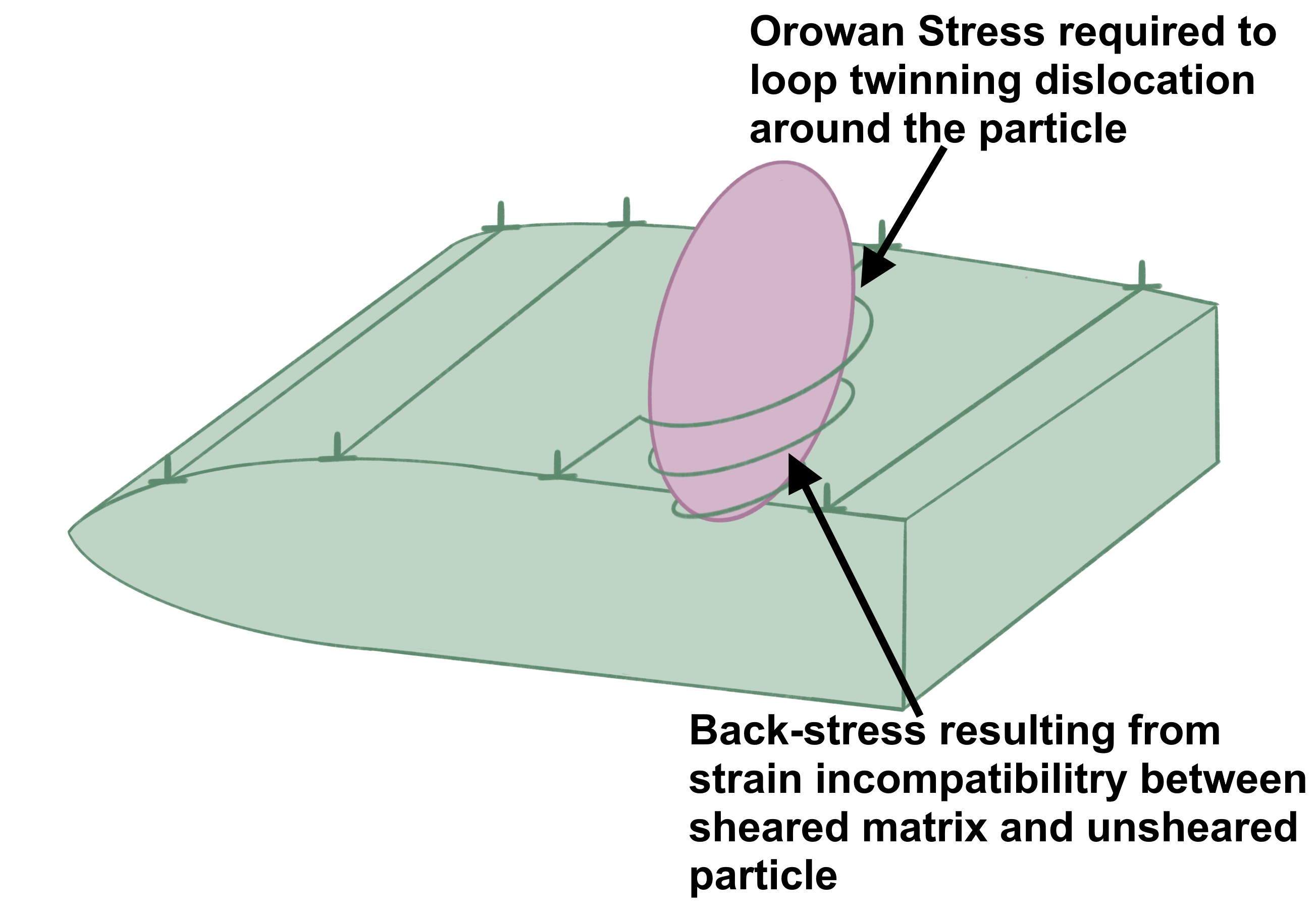}
    \caption{Interaction between twinning dislocations and a precipitate.}
    \label{fig:slipinteraction}
\end{figure}

\begin{figure}[h]
    \centering
    \includegraphics[width=0.85\linewidth]{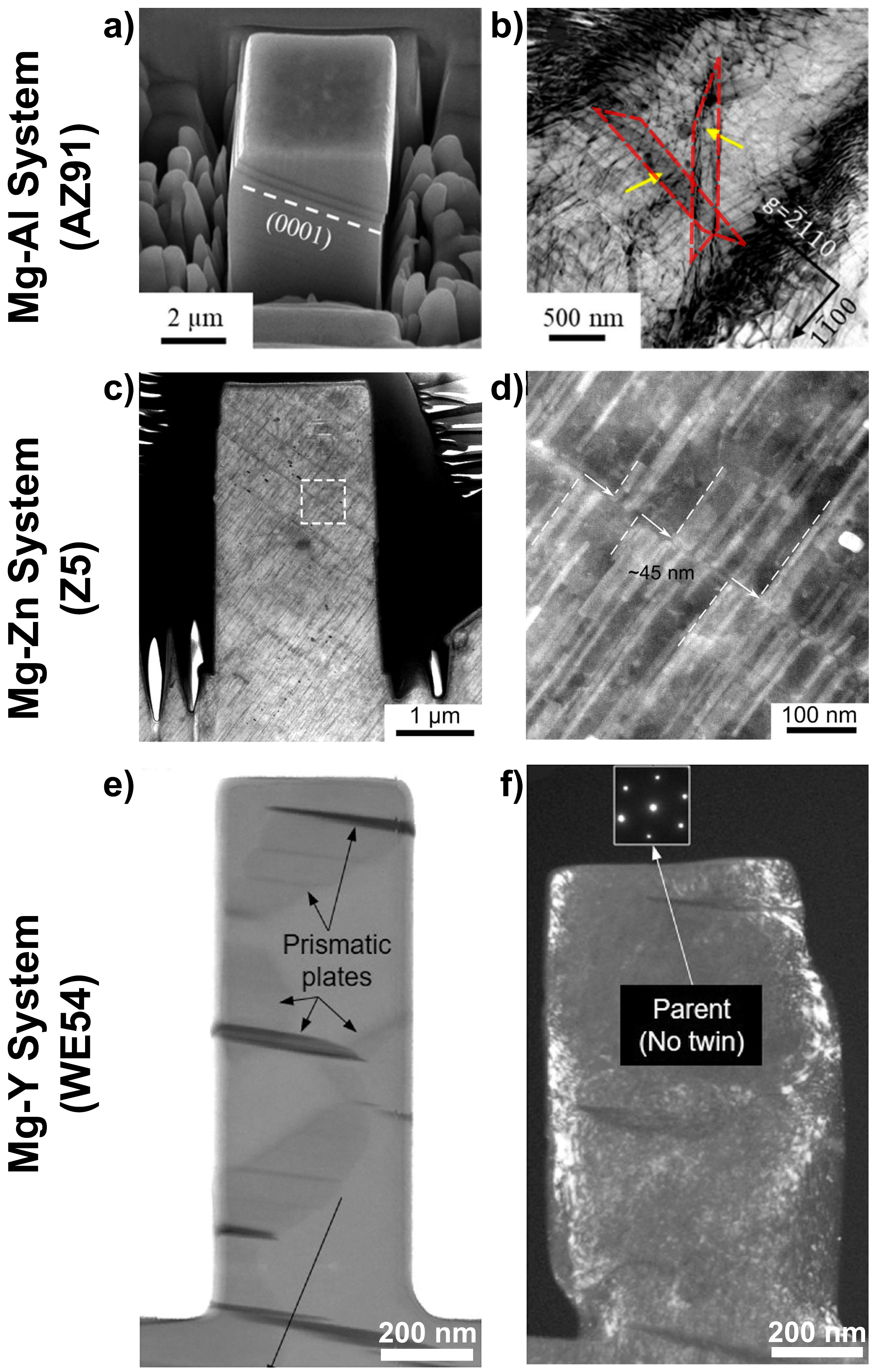}
    \caption{Micropillar compression images featuring Mg alloys showing examples of basal plates, $c$-axis rods, and prismatic plate systems. (a,b) \cite{Si2020}  (permission request being processed) (c,d) \cite{Wang2015} Reprinted from Acta Materialia, Vol 100, Jiangting Wang, Nicole Stanford, Investigation of precipitate hardening of slip and twinning in Mg5\%Zn by micropillar compression, Pages 53-63, Copyright 2015, with permission from Elsevier.  (e,f) \cite{Liu2018Insight} Reprinted from Journal of Materials Science \& Technology, Vol 34 Issue 7, Bo-Yu Liu , Nan Yang, Jian Wang , Matthew Barnett, Yun-Chang Xin, Di Wu, Ren-Long,Xin ,Bin Li ,R. Lakshmi Narayanag, Jian-Feng Nie , Ju Li ,Evan Ma, Zhi-WeiShan, Insight from in situ microscopy into which precipitate morphology can enable high strength in Magnesium alloys, Pages 1061-1066, Copyright 2018, with permission from Elsevier.}
    \label{fig:micropillarTEM}
\end{figure}

The impenetrable precipitates have two main contributions to the strengthening against twinning: Orowan stress and back stress as shown in \cref{fig:slipinteraction}. The Orowan stress arises when the twinning dislocations loop around the particle, and the back stress results from strain incompatibility between the sheared matrix and unsheared particle. The plate-shaped precipitates inhibit tensile twin growth by introducing incompatibility stresses in the twinned region. A recent MD simulation study by \cite{RN2116} suggests that precipitates provide stronger resistance to twin thickening than to dislocation slip. Effects of the shape and orientation of the precipitates were also examined in a phase-field study \citep{RN2118}. \cite{RN2117} used a dislocation-based twin propagation model to demonstrate that an Orowan strengthening mechanism is still valid for explaining how precipitates inhibit twin propagation. In addition, twinning was found to rotate the precipitates in the parent matrix only slightly.  As a result, the basal plates, which originally provide weak resistance to basal slip, become much stronger obstacles to twinning \citep{robson2013effect}. Micro-pillar compression studies allow testing of single crystal Mg alloys with different precipitate types, as shown in \cref{fig:micropillarTEM}. These studies are helpful in obtaining CRSS values as well as understanding the role of twin-precipitate interactions. \\

\subsection{Interaction of solute atoms, solute clusters and precipitates with grain boundaries and texture }
\label{subsec:SoluteTexture}
\begin{figure}[h]
    \centering
    \includegraphics[width=\linewidth]{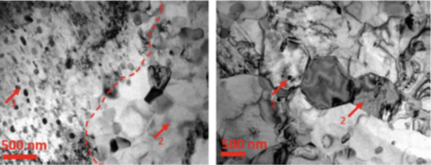}
    \caption{Nanoprecipitates being consumed by an advancing combined reaction zone in Mg-9Al (wt\%) alloy being deformed by the ECAE (equal channel angular extrusion) process. Figure from \cite{21}. Reprinted by permission from Springer Nature Customer Service Centre GmbH: Springer Nature Magnesium Technology 2020 , Effect of Second Phase Particle Size on the Recrystallized Microstructure of Mg–Al Alloys Following ECAE Processing, Suhas Eswarappa Prameela, Peng Yi, Vance Liu, Beatriz Medeiros, Laszlo J. Kecskes, Michael L. Falk, Timothy P. Weihs, 2020 }
    \label{fig:ECAETEM}
\end{figure}

There are three main mechanisms through which solute atoms, solute clusters, and small precipitates influence grain boundaries and the resulting texture in Mg alloys. First, solute atoms can segregate to grain boundaries where they may form clusters and prevent grain growth and rotation.  For example, in Mg-Zn-Ca alloys, fine intermetallic particles can restrict grain boundary motion and rotation via Zener pinning \citep{hofstetter2015processing}. Thus, one can strategically control segregation, clustering, and precipitation to indirectly aid in grain boundary strengthening. Second, by selectively adding elements, one can promote non-basal deformation modes in Mg alloys. Studies have  shown that adding certain RE additions produce what is called “RE texture” \citep{robson2013reducing}. The addition of RE elements creates greater accommodation of both compressive and tensile deformation, leading to unique texture changes that are otherwise seldom seen in Mg alloys. The third way solute atoms influence grain boundaries is through solute decomposition and precipitation along grain boundaries. Precipitates can form readily along grain boundaries during processing. Recrystallization, another strengthening reaction, typically occurs in the same regions. These two reactions can interact with each other, producing combined reaction zones. The combined reaction zone is a mix of intermetallic particles and $\alpha$-Mg matrix grains. The spread of this zone is heavily dependent on solute concentrations in the matrix and the precipitates lining the boundaries. For example, \citep{21} have shown that, with more precipitation within the grain interiors, the size of the combined reaction zone may also be increased, as shown in \cref{fig:ECAETEM}. Thus, both solute clusters and precipitates influence microstructural evolution, which in turn affects mechanical properties.

\subsection{Influence of solute atoms, solute clusters and precipitates on global yield response }
\label{subsec:SoluteYield}
\vspace*{1mm} 
\begin{figure}[]
    \centering
    \includegraphics[width=\linewidth]{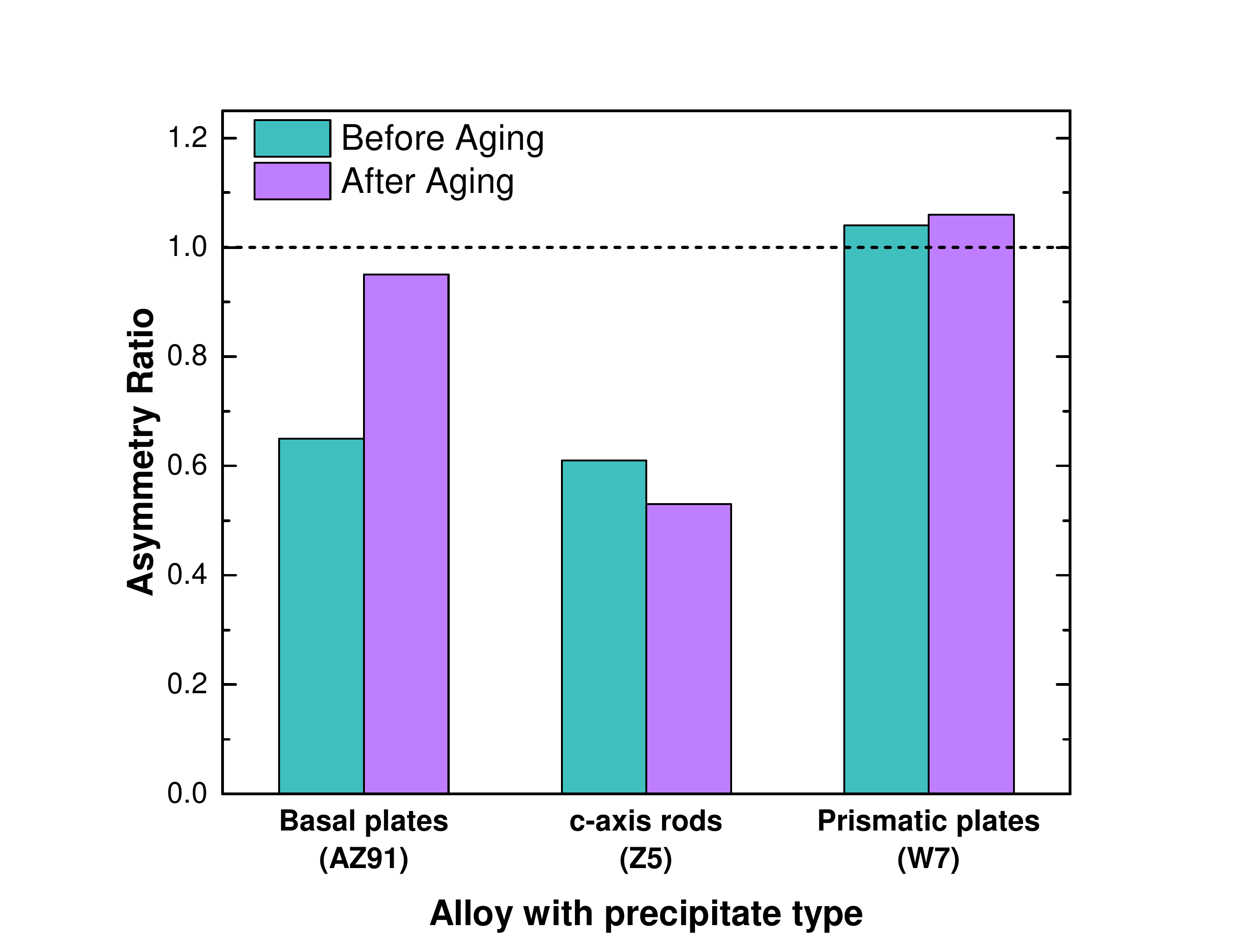}
    \caption{Asymmetry ratio ($\sigma_{yield}^{compression}/\sigma_{yield}^{tension}$) of assorted Mg alloys before and after peak aging. Data from \cite{davis2019reducing}.}
    \label{fig:asymmetryratios}
\end{figure}

Here we characterize the cumulative impact of the interactions discussed in \cref{subsec:SoluteDislocation,subsec:SoluteTwin,subsec:SoluteTexture} on the material's overall yield response. In many Mg alloys, which have a low concentration of solute atoms, solute clusters, or precipitates in the matrix, result in a minimal impact on the global yield response. In other alloys, a small fraction of clusters or precipitates at specific locations (such as grain boundaries) can dramatically change the yield response. Therefore, it is essential to assess what volume fraction of precipitates is needed to significantly change the yield response. \cite{robson2013reducing} showed that for basal plates, a significant volume fraction ($>$10\%) and texture weakening are both important to improve the strength and reduce the tension-compression yield asymmetry observed in Mg alloys. Precipitates also help tailor the ductility and the strain rate sensitivity of the material. Studies have shown that a significant volume fraction of precipitates alters the yield asymmetry (ratio of yield strength in compression to tension)  \citep{davis2019reducing}. \cref{fig:asymmetryratios} shows how the asymmetry ratio changes before and after aging for alloys with basal plates, $c$-axis rods, and prismatic plates. While basal plates are shown to reduce the asymmetry, $c$-axis rods typically increase the asymmetry due to their orientation. The prismatic plates tend to have a minimal effect on yield asymmetry. 

%\begin{table}[h]
%\centering
%\label{asymmetryratios}
%\caption{Asymmetry ratio ($\sigma_{yield}^{compression}/\sigma_{yield}^{tension}$) of assorted Magnesium alloys before and after heat treatment. Data from \cite{davis2019reducing}.}
%\begin{tabular}{|l|l|l|}
%\hline
%Alloy & Before-aging AR & After-aging AR \\ \hline
%AZ91 (basal plate) & 0.65 & 0.95 \\ \hline
%TZX620 (basal plate) & 0.46 & 0.53 \\ \hline
%Z5 (c-axis rods) & 0.61 & 0.53 \\ \hline
%W7 (prismatic plate) & 1.04 & 1.06 \\ \hline
%\end{tabular}
%\end{table}
\subsection{Modeling of deformation mechanisms in Mg alloys}

Crystal plasticity models have been adopted to model the interplay of Mg's slip and twin systems. Unlike dislocation slip, which is well understood and captured in crystal plasticity and phase field models, modeling deformation twinning still presents challenges, and a variety of models exist for slip and twinning in HCP metals. A simple way to include twinning is the introduction of additional slip systems (often referred to as \emph{pseudo slip systems})~\citep{VanHoutte1978, Kalidindi1998, AgnewEtAl2001, StaroselskyAnand2003, GraffEtAl2007}, so that the local average twin volume fraction becomes the evolving internal variable on the mesoscale, within an otherwise classical crystal plasticity-based formulation at finite strains, cf.~\citep{rice1971inelastic,AsaroRice1977,Asaro1983}.  \cite{walde2007} used the approach of \cite{tome1991model} to model twinning, which relies on a Schmid-type formulation for the resolved shear stress on each twin system. 
% The theoretical twin shear is computed for each system, while only the most active system--the predominant twin system (PTS)--is activated. In this way the activation of multiple twin systems within a grain is prohibited; the twin evolution is computed solely for the PTS. The effect of twins on the slip systems is modeled through the evolution of the twin volume fractions and a fixed crystallographic relationship is maintained from parent to twin, once the twin volume fraction exceeds a threshold. Based on \cite{tome1991model} and \cite{tome2005modeling}, \cite{proust2007modeling} developed the composite grain (CG) model (for Zr), which captures the formation of twins more accurately, in the sense that multiple twin lamellae may form across a grain, as observed in experiments. In the spirit of the PTS approach, only the twin system with the largest accumulated volume fraction forms lamellae within a grain; new lamellae are formed such that the centerline of the twins are equidistant (see \cref{fig:compositeGrain}). By effectively cutting the parent phase of the grain into multiple, smaller regions, the twin lamellae act as barriers to slip, thus introducing an orientation-dependent Hall-Petch type hardening. 
Unlike most prior approaches, 
\cite{zhang2012phenomenological} developed a model for pure Mg that accounts for the crystal reorientation due to twinning. %Their results captured the experimental findings of \cite{kelley1968plane} accurately and predicted texture evolution in polycrystalline Mg. This served as the basis for the refined approaches of \cite{indurkar2020predicting}, predicting the texture dependence of plastic hardening. Also, validated by the data of \cite{kelley1968plane}, 
\cite{chang2015variational} used a similar formulation based on twin volume fractions to model plasticity in Mg, but adopted a variational setting and showed advantages in terms of, for instance, the achievable time steps and the stability of the numerical scheme. Also based on a variational structure, \cite{homayonifar2010} introduced a model for Mg that models deformation twinning as a phase transformation process and exploits lamination by energy relaxation. \\

Reduced-order models sacrifice microstructural details for computational efficiency. \cite{becker2016reduced} presented a reduced-order HCP crystal plasticity model, which accounts only for basal and twin systems explicitly, while lumping all other slip systems into an effective yield-surface-type description of plasticity. Since this model neglects the lattice reorientation due to twinning, it is most accurate at small to moderate strain levels, before twin saturation is reached. The obvious benefit of this model is reduced complexity and computational costs when compared to higher-fidelity crystal plasticity models.\\

% \begin{figure}
%     \centering
%     \includegraphics[width= \linewidth]{Figures/Proust_composite_grain.png}
%     \caption{Left: EBSD data of Zr deformed under through-thickness compression at 5\%. Middle: Schematic of the CG grain model. Twin lamellae are shown in yellow, parent grain in green. Right: Uncoupled representation of the matrix and twin regions to show the characteristic lengths of the model. Taken from \cite{proust2007modeling}. (permission request being processed)}
%     \label{fig:compositeGrain}
% \end{figure}

\begin{figure}[htb]
\includegraphics[width= \linewidth]{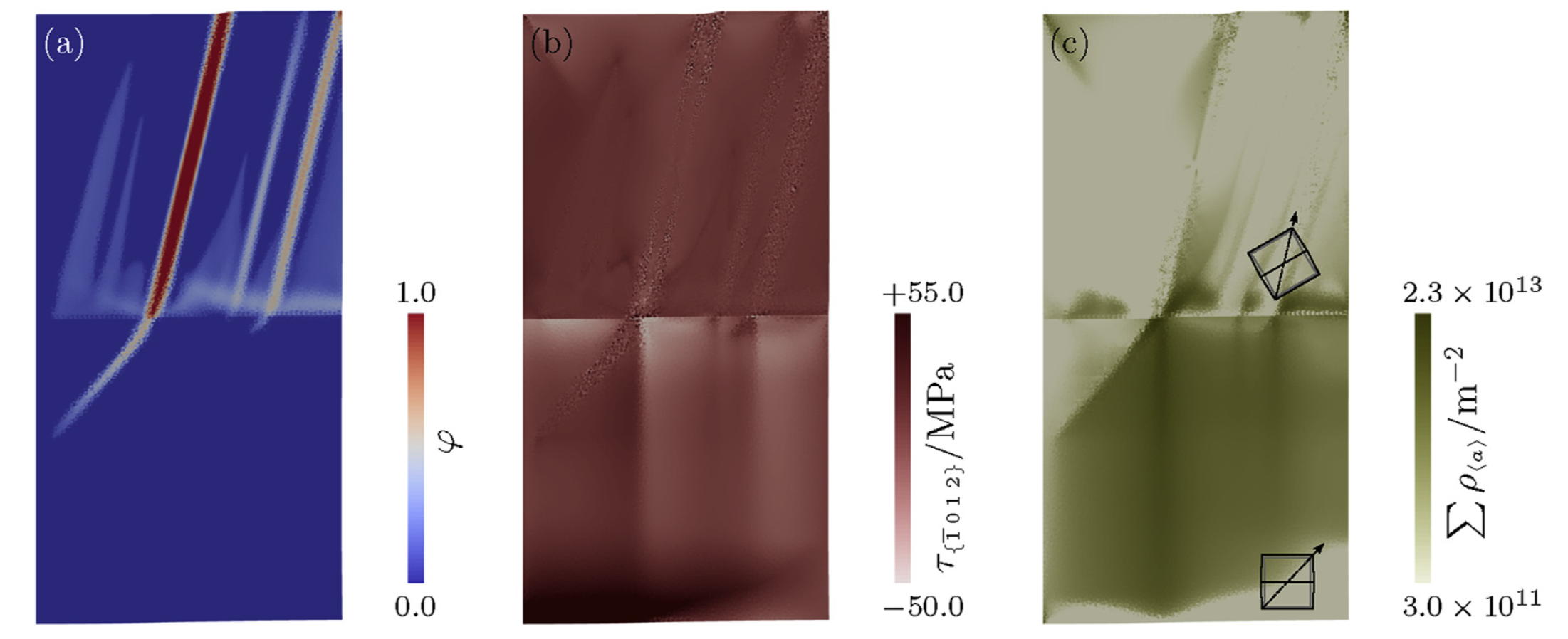}
\caption{Evolution of the twin volume fraction, the resolved shear stress, and the basal dislocation densities across the interface of a pure Mg bicrystal, as simulated by \cite{liu2018integrated}. Reprinted from International Journal of Plasticity, Vol 106, C. Liu,P. Shanthraj,M. Diehl,F. Roters,S. Dong,J. Dong,W. Ding,D. Raabe, An integrated crystal plasticity–phase field model for spatially resolved twin nucleation, propagation, and growth in hexagonal materials, 203-227, Copyright 2018, with permission from Elsevier. }
\label{fig:LiuBicrystal}
\end{figure}

% While variational models are elegant and offer rigorous mathematical formulations, they are limited in the sense that it can prove hard (or impossible) to add more detailed physics while preserving the variational formulation. Therefore, nonvariational approaches often trade mathematical elegance for more involved physical descriptions. 

While the above crystal plasticity approaches are efficient and accurate for predicting the effective response of Mg and its alloys (if appropriately calibrated), they include limited physical insight into microstructural mechanisms. This latter insight is the goal of more refined models such as the one by \cite{liu2018integrated}, which describes the evolution of dislocation densities on all slip systems while solving a phase-field problem for twinning (see \cref{fig:LiuBicrystal}).\\ %Models of the evolution of dislocation densities that also capture the intricate interactions between slip and twin systems and material size effects provide powerful insight towards understanding and predicting the formation of precipitate clusters based on, among others, local slip activity. \\

In the materials by design approach, we aim to obtain critical information about the link between process conditions and the resulting microstructures of Mg alloys. Consider a Mg alloy undergoing a thermomechanical processing such as rolling or extrusion. In this case, both deformation and temperature are being simultaneously imparted to the sample. The microstructure continuously evolves, and we often obtain information about the microstructure as well as the presence and distribution of precipitates after processing. By feeding the initial texture data into a polycrystal simulation, we can simulate the texture evolution and deformation mechanisms in the polycrystalline sample during the processing step. We can then attempt to identify connections between the local state of deformation as predicted by simulations and the observed precipitate distribution in experiments, with the goal of identifying causal mechanisms for precipitation. This can enrich the model by adding information about, for example, grain size effects \citep{ravaji2021crystal}, precipitate-, solute- and solute cluster hardening. \\

The precipitation model by \cite{hutchinson2005modeling} includes several typical hardening mechanisms. Notably, the authors consider the effects of grain size, solute strengthening, and precipitate strengthening in their model formulation.
Following \citep{caceres2001solid}, solute strengthening can be modelled by 
\begin{equation}
    \sigma_{\text{SS}} = CX^{2/3},
\end{equation}
whose parameter $C$ in Mg-Al alloys was identified as 

$197 (MPa X^{-2/3})$, with $X$ being the atomic fraction of solute in the alloy. The results of \cite{caceres2001solid} are in agreement with the experimental data of \cite{Akhtar1972}. When it comes to precipitate strengthening, \cite{hutchinson2005modeling} computed the shear resistance of the precipitate inclusion and found the critical precipitate radius above which dislocations bow around the particles to be at $\approx 3nm$. Hence, they assumed that it is safe to ignore shearing of precipitates, as the critical radius for nucleation is on the same order as the critical radius for shearing versus bowing \citep{hutchinson2005modeling}.
The precipitate hardening effects are thus computed in a similar fashion to \cref{OrowanEquation}; for more details, see e.g. \cite{brown1971strengthening}.\\
% \begin{equation}
%     \sigma_{\text{Orowan}} = \frac{MG_m\bm{b}}{2\pi\sqrt{1-\nu}} \frac{1}{\lambda}\left(\frac{d_A}{r_0} \right)
% \end{equation}
% with Taylor factor $M$, shear modulus $G_m$, Poisson's ratio $\nu$, Burgers vector $\bm{b}$, and the average planar interparticle spacing $\lambda$. $d_A$ stands for the mean diameter of the particle in the slip plane and $r_0$ is the inner cut-off radius of the dislocation, taken equal to $\bm{b}$ \citep{hutchinson2005modeling}.\\

Despite much progress on modeling precipitation in Mg alloys (including the above approaches and those mentioned in \cref{subsec:Precipitationmodeling}), an accurate model that predicts the simultaneous evolution of texture and precipitation in Mg alloys during severe plastic deformation processing is an open challenge and requires both improved mesoscale plasticity models and detailed insight into the microscale processes leading to precipitate nucleation and growth and these  processes' interactions with crystal defects.

\section{Controlling the clustering and nucleation process}
\label{sec:ControllingClustering}

\subsection{Experimental observations} 
\label{subsec:ExperimentalObservations}
\vspace*{1mm}

\begin{figure}[h]
    \centering
    \includegraphics[width=0.85\linewidth]{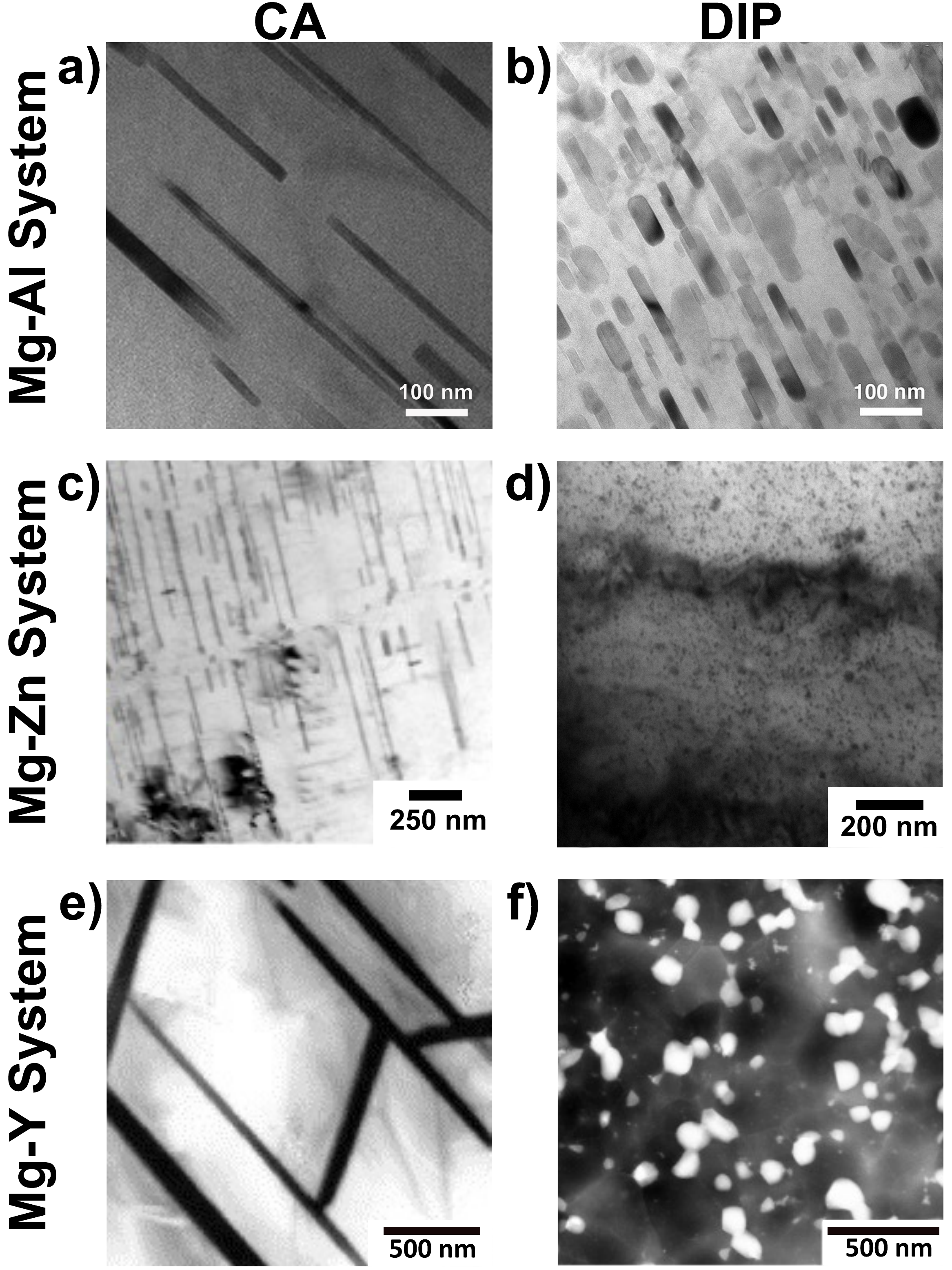}
    \caption{TEM micrographs of precipitates from conventionally aged and DIP processed Mg alloys. (a) Basal plates from conventionally peak aged A9 alloy \citep{ma2019dynamic} \protect \footnotemark[1] (b) Precipitates in hot-ECAE processed A9 alloy \citep{ma2019dynamic} (c) $c$-axis rods in conventionally aged Z6 alloy\citep{psref4} \protect \footnotemark[2] (d) Precipitates in hot-ECAE processed Z3 alloy \citep{prameela2020enhanced}  \protect \footnotemark[3] (e) Prismatic plate precipitates in conventionally aged Y10 alloy\citep{Zhang2003} \protect \footnotemark[4](f) Prismatic plate precipitates in hot-ECAE processed WE43 alloy \citep{Minrik2018} \protect \footnotemark[5]}
    \label{fig:PrecipitateTEMs}
\end{figure}

\footnotetext[1]{Reprinted from Acta Materialia, Vol 172, X.L.Ma, Suhas Eswarappa Prameela, Peng Yi, Matthew Fernandez, Nicholas M.Krywopusk, Laszlo J. Kecskes, Tomoko Sano, Michael L.Falk, Timothy P.Weihs, Dynamic precipitation and recrystallization in Mg-9wt\%Al during equal-channel angular extrusion: A comparative study to conventional aging, Pages 185-199, Copyright 2019, with permission from Elsevier.}
\footnotetext[2]{Reprinted from Scripta Materiala, Vol 60 Issue 6,  Hidetoshi Somekawa, Alok Singh, Toshiji Mukai, Microstructure evolution of Mg–Zn binary alloy during a direct extrusion process, Pages 411-414, Copyright 2009, with permission from Elsevier.}
\footnotetext[3]{Reprinted from Procedia Manufacturing, Vol 47, Suhas Eswarappa Prameela,Elaine Lipkin,Joey Chen,Laszlo Kecskes,Zhigang Xu, Timothy Weihs, Enhanced Precipitation and Recrystallization in a Mg-Zn Alloy During Low-Temperature Extrusion, Pages 1476-1480, Copyright 2020, with permission from Elsevier.}
\footnotetext[4]{ Reprinted from Scripta Materialia, Vol 48 Issue 4, M.-X. Zhang, P.M. Kelly, Morphology and crystallography of Mg24Y5 precipitate in Mg–Y alloy, Pages 379-384, Copyright 2003, with permission from Elsevier. }
\footnotetext[5]{Reprinted from Materials Characterization, Vol 140, Peter Minárik, Jozef Veselý, Jakub Čížek, Mária Zemková, Tomáš Vlasák, Tomáš Krajňák, Jiří Kubásek, Robert Král,Daniel Hofman, Jitka Stráská, Effect of secondary phase particles on thermal stability of ultra-fine grained Mg-4Y-3RE alloy prepared by equal channel angular pressing, Pages 207-216, Copyright 2018, with permission from Elsevier.}

Understanding the strengthening mechanisms as discussed in \cref{sec:EffectSolute} provides guidance on how one can improve the strengthening effects in Mg alloys, which are otherwise weak compared to other metals (\cref{fig:AgingStrengths}).  This understanding also helps us examine how one one can use deformation to control and enhance the strengthening effects of these three microstructure features (solutes, clusters, and precipitates).  In the remainder of this review, we ignore grain size strengthening in order to focus primarily on precipitate strengthening and offer some insights regarding solute and solute cluster strengthening.  \cref{fig:PrecipitateTEMs} (a)(c)(e) shows the typical precipitate morphology in Mg alloys that are prepared using conventional aging (CA) conditions, which involve quenching a sample solutionized at high temperature and then maintaining a constant temperature. These precipitates often take the shape of laths with very high aspect ratios and are spaced far apart.  Based on the Orowan mechanism (see \cref{sec:EffectSolute}), these precipitates are relatively ineffective at blocking dislocations and twins in Mg alloys, resulting in very limited precipitate strengthening.\\

An effective way to overcome this limited strengthening is to introduce deformation before or during thermal processing to reduce the size and spacing of the precipitates and to improve their morphology.  We term this method Deformation-Induced Precipitation (DIP).  Several studies in the last decade have shown that the DIP process can produce Mg alloys with finer precipitates than conventional thermal heat treatments.  For example, \cref{fig:PrecipitateTEMs} (b)(d)(f) shows that precipitates produced through DIP are far smaller, more finely spaced, and more equiaxed than precipitates produced in the same alloys via conventional aging.  Due to these changes, the three alloys have displayed improved strengthening compared to conventional aging \citep{prameela2020deformation,ma2019dynamic,prameela2020enhanced}, as predicted by the precipitate strengthening model discussed in \cref{sec:EffectSolute}.\\

\begin{figure}[h]
    \centering
    \includegraphics[width=\linewidth]{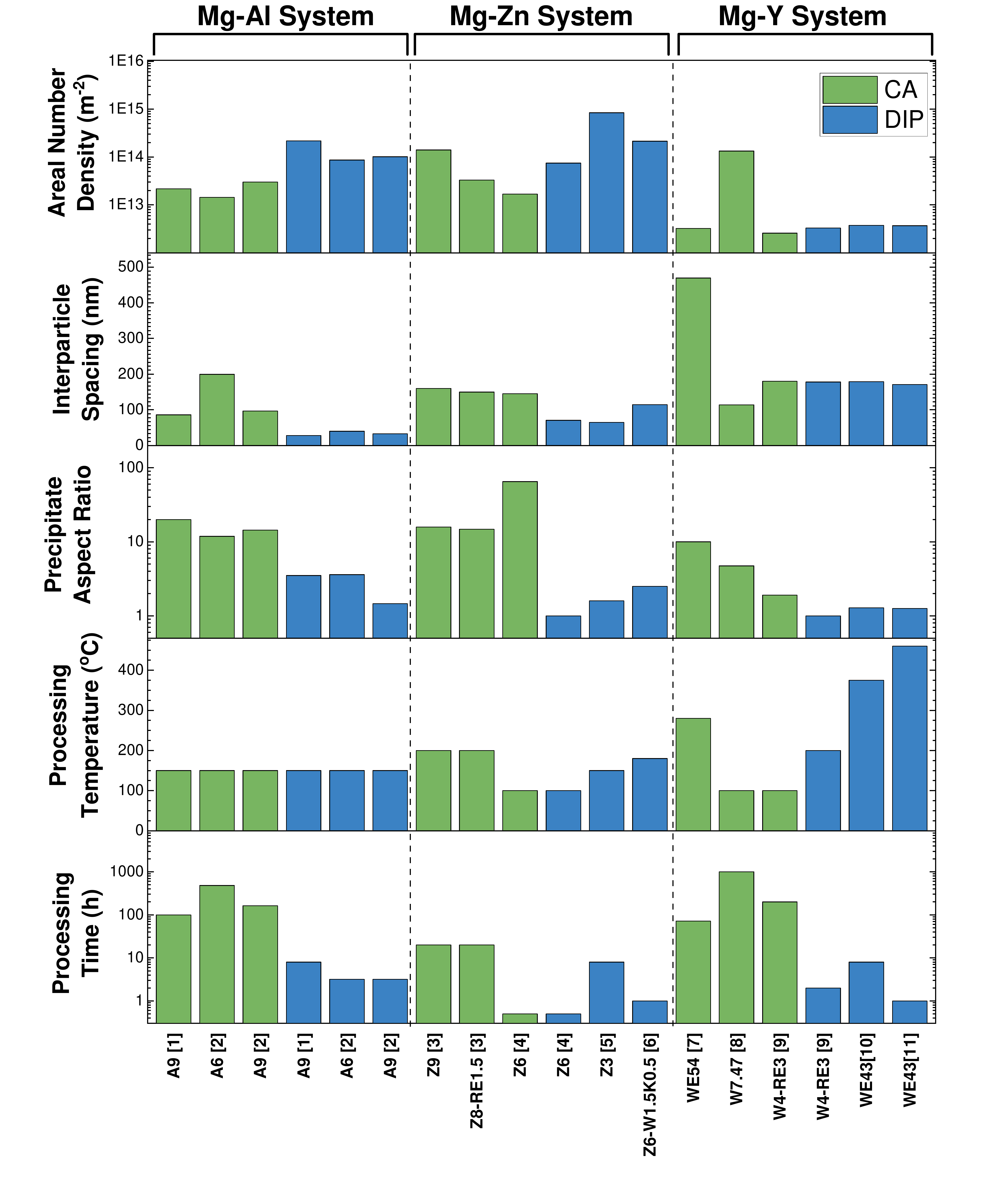}
    \caption{ Precipitate and processing data for Mg-Al, Mg-Zn, and Mg-Y alloy systems (both CA and DIP). Data from sources as indicated: [1] \cite{ma2019dynamic} [2] \cite{prameela2020deformation} [3] \cite{psref3} [4] \cite{psref4} [5] \cite{prameela2020enhanced} [6] \cite{psref6} [7] \cite{psref7} [8] \cite{psref8} [9] \cite{psref9} [10] \cite{kang2015achieving} [11] \cite{cao2015superplastic} }
    \label{fig:PrecipStats}
\end{figure}

A comparison of precipitate characteristics such as aspect ratio, inter-particle spacing, and areal number density between conventional processing and DIP processing of Mg-Al, Mg-Zn, and Mg-Y alloys is shown in \cref{fig:PrecipStats}.  These numbers indicate that DIP enhances the nucleation of precipitates and provides a high areal number density, a smaller interparticle spacing, and a lower aspect ratio. These parameters in turn lead to higher levels of strengthening.\\

\begin{figure}
    \centering
    \includegraphics[width=0.8\linewidth]{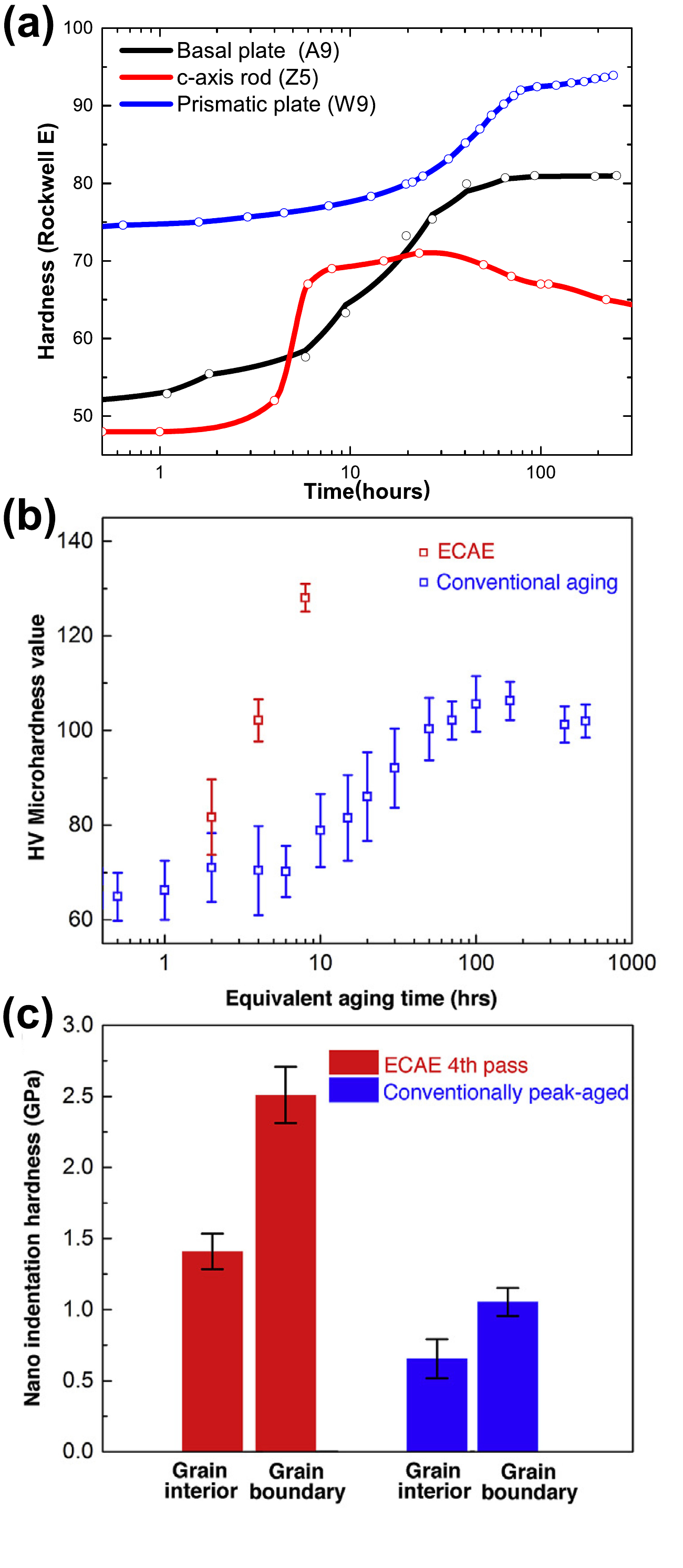}
    \caption{(a) Conventional aging curves at 200 $^{\circ}$C for A9 \citep{clark1968age}, Z5 \citep{clark1965transmission}, and W9 \citep{mizer1972study} binary alloys. Cutoffs at reported solubility limit. (b) Microhardness measurements of conventionally aged and dynamically processed samples as a function of time at 150°C. (c) Site-specific nanoindentation hardness in grain interiors and at grain boundaries in samples after four B$_c$ ECAE passes and conventionally peak-aged samples. Each error bar is the standard deviation resulting from 10 indentation measurements. (b) and (c) from \cite{ma2019dynamic}. Reprinted from Acta Materialia, Vol 172, X.L. Ma, Suhas Eswarappa Prameela, Peng Yi, Matthew Fernandez, Nicholas M. Krywopusk, Laszlo J. Kecskes, Tomoko Sano, Michael L. Falk, Timothy P. Weihs, Dynamic precipitation and recrystallization in Mg-9wt.\%Al during equal-channel angular extrusion: A comparative study to conventional aging, Pages 185-199, Copyright 2019, with permission from Elsevier.}%\protect\footnotemark[6].}
    \label{fig:AgingCurves}
    \label{fig:DIPagingcurves}
\end{figure}

The enhanced nucleation shown in \cref{fig:PrecipitateTEMs} and \cref{fig:PrecipStats} is also accomplished in a much shorter time than conventional aging, as noted in \cref{fig:PrecipStats}.  The time to peak aging after deformation can be up to 100x shorter than for conventional aging at the same process temperatures, which reduces cost and energy use.  A typical response to a conventional thermal aging treatment of a Mg alloy is shown in \cref{fig:AgingCurves}, which plots the samples' hardness as a function of heat treatment or aging time. The initial increase in hardness is due to a rise in the number density and size of precipitates via nucleation and growth. The eventual decrease after the peak is due to coarsening of the precipitates at a fixed volume percentage, which reduces their number density, thus also reducing their strengthening.  The time to reach maximum hardness is called the peak aging time.  There are two main disadvantages of Mg alloys compared to Al alloys. First, the peak aging times for Mg alloys are very long. Secondly, the peak hardness increase is much lower ($\sim2\times$) than the enhancements we typically see in Al alloys. Both of these disadvantages can be significantly alleviated using DIP, as shown in \cref{fig:DIPagingcurves}, due to the enhanced nucleation of precipitates.\\

%\begin{figure}
%    \centering
%    \includegraphics[width=\linewidth]{Figures/Solubility vs Temp ver2.pdf}
%    \caption{Solubility Curves for Al, Zn, and Y in Mg solid solution}
%    \label{fig:SolubilityVsT}
%\end{figure}

%\begin{figure}[h]
%    \centering
%    \includegraphics[width=\linewidth]{figure 3.1.png}
%    \caption{Standard aging treatments for Mg alloys (AZ91, Z5, Z8) where the hardness reaches the peak aging condition and then drops off. Figure from \cite{22}}
%    \label{fig:AgingCurves}
%\end{figure}
%we're taking this figure out

%\footnotetext[6]{Reprinted from Acta Materialia, Vol 172, X.L. Ma, Suhas Eswarappa Prameela, Peng Yi, Matthew Fernandez, Nicholas M. Krywopusk, Laszlo J. Kecskes, Tomoko Sano, Michael L. Falk, Timothy P. Weihs, Dynamic precipitation and recrystallization in Mg-9wt.\%Al during equal-channel angular extrusion: A comparative study to conventional aging, Pages 185-199, Copyright 2019, with permission from Elsevier }

\subsection{Basics of nucleation of precipitates}
\label{hetnucleation}
\vspace*{1mm}

To understand the higher rate of precipitate nucleation during DIP, we turn to the classical nucleation theory (CNT) that is reviewed extensively in %\textcolor{red}{[Christian 1975 textbook]}
\cite{christian1975textbook}.  In brief, the nucleation rate can be expressed as 
\begin{equation}
    I=I_0 \exp(-\frac{\Delta G^*}{k_b T}),
\end{equation}
where $\Delta G^*$ is the change in free energy required to form a precipitate of the critical size in a matrix and, hence, the energy fluctuation required to overcome the nucleation barrier. This free energy barrier to nucleation is given by
\begin{equation}
    \Delta G= V (\Delta g_\nu + \Delta g_s) + A \sigma,
\end{equation}
where $\Delta g_v$ is the free energy difference between the precipitating phase and the matrix per unit volume; $\Delta g_s$ is the misfit elastic strain energy of the precipitate nucleus in the matrix; $\sigma$ is the interfacial energy, and $V$ and $A$ are the volume and surface area of the nucleus respectively. The chemical driving force $\Delta g_v$ can often be extracted from accurate first-principles calculations.  The elastic strain energy $\Delta g_s$ for a coherent precipitate nucleus is given by
\begin{equation}
    \Delta g_s = \frac{2G_m(1+\nu)}{1-\nu} \epsilon^2,
\end{equation}
%\todo{conflicting abbreviation, G}
where $G_m$ is the shear modulus, $\nu$ is Poisson’s ratio, and $\epsilon$ is the misfit parameter defined as $(v_p - v_m) / 3v_m$, where $v_p$ and $v_m$ are average atomic volumes of the precipitate and the matrix, respectively.  The interfacial free energy is the hardest to determine, and its value varies significantly ($\sim 0.1-1 J/m^2$) depending on whether the interface is coherent, semi-coherent, or incoherent.\\

Due to the large concentration difference between the initial solid solution and the precipitate phase, the barrier to nucleation, $\Delta G^*$, is high in Mg.  Furthermore, the diffusion of solute atoms in Mg is low compared to Al. Both these factors lead to low nucleation rates and a low number density of precipitates in Mg during conventional aging. Next, we explore how one can introduce defects through deformation during processing to overcome these two limitations and enable higher nucleation rates and more effective strengthening in Mg alloys.\\

%\begin{figure}[]
%    \centering
%    \includegraphics[width=\linewidth]{Figures/nucleation free energy ver3.pdf}
%    \caption{Schematic of free energy curves of Homogeneous and Heterogeneous nucleation and growth.}
%    \label{fig:nucleationenergy}
%\end{figure}

Heterogeneous nucleation is far more common than homogeneous nucleation in real materials because of impurities, defects, and boundaries that serve as nucleation sites \citep{Porter-Easterling}. When desired, one can even artificially introduce nucleation sites to boost the nucleation rate by reducing the interfacial free energy or increasing the driving force.  For example, it is common to introduce seed particles during the solidification process.  These seed particles reduce the interfacial free energy cost, enhance the nucleation of intermetallics, and lead to fine grain sizes~\citep{Jiang2018,Li2020}.\\

\subsection{Clustering and heterogeneous nucleation via deformation induced defects}
\vspace*{1mm}

One viable way to boost nucleation kinetics is through DIP at low temperatures, as mentioned in \cref{subsec:ExperimentalObservations}. When polycrystalline samples with multiple grains are deformed, the load distribution results in varying levels of strain within the grains and along grain boundaries. The plastic deformation results in the creation of several types of defects (vacancies, dislocations, stacking faults, twins). The propensity to form these defects and the defects' density are dictated by the temperature, amount of deformation, and activation of specific slip and twinning systems. The key idea behind DIP is that by controlling the motion, density, and heterogeneity of defects generated through thermo-mechanical treatments, the solutes and solute clouds resulting from these defects can be utilized to form desired precipitate microstructures, such as those illustrated in \cref{fig:PrecipitateTEMs}. It is crucial to understand both how these defects influence the formation of solute clusters or precipitates and how the defect density changes during deformation in order to maximize strength via DIP methods.  Below we focus on two types of defects: dislocations and vacancies.\\

\subsubsection{Mechanisms of dislocation induced clustering and nucleation}
\label{dislocationhetnucleation}
\vspace*{1mm} 

As \cite{Porter-Easterling} summarized, dislocations can assist precipitate nucleation kinetically by providing solute diffusion ``pipes", and thermodynamically by inducing the segregation of solute atoms.  The pipe diffusion mechanism can boost the diffusivity of solute atoms by up to 3 orders of magnitude \citep{Legros2008}, but the latter's effect is more significant, as will be discussed below.\\

\begin{figure*}[h]
    \centering
    \includegraphics[width=0.95\linewidth]{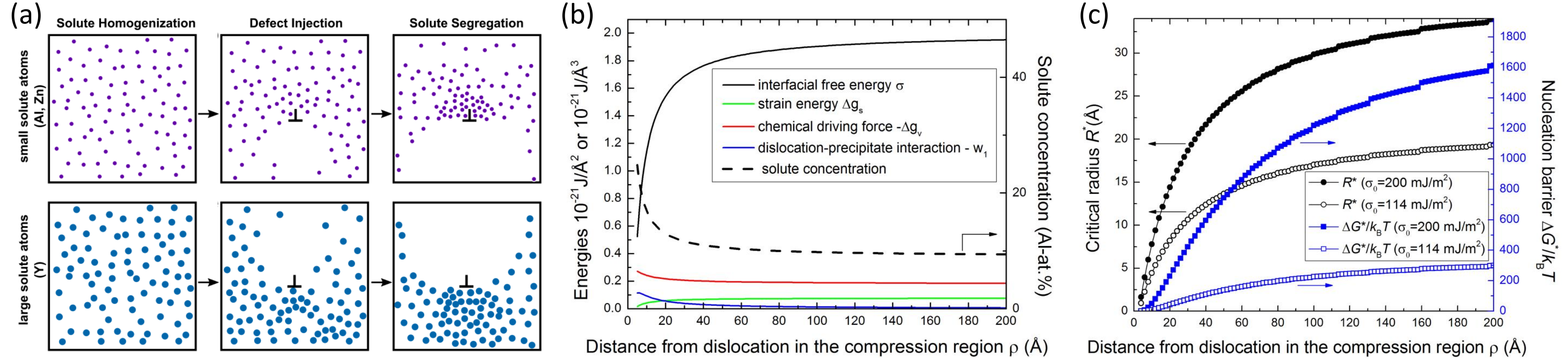}
    \caption{(a) Solute atom segregation around an edge dislocation. Solute atoms larger than Mg tend to segregate in the bottom compression region and smaller atoms tend to segregate to the top of the dislocation core. (b) The interfacial free energy $\sigma$, the chemical driving force $\Delta g_v$, the misfit strain energy $\Delta g_s$, and the dislocation effect $w_1$, as functions of $y$, distance from the dislocation core in the vertical direction. (c) The critical radius $R^*$ and the nucleation barrier $\Delta G^*$ as functions of $y$. Figure from \citep{ma2019dynamic}.\protect\footnotemark[7]}
    \label{fig:SoluteDislocationSegregation}
\end{figure*}

Solute atoms segregate near edge dislocations to reduce the elastic strain energy in the system, forming a solute “cloud” or Cottrell atmosphere, as seen in \cref{fig:SoluteDislocationSegregation}(a). The Cottrell atmosphere can be predicted by the Cottrell-Eshelby (CE) model \citep{37,38}, but the CE model does not consider the effect of the segregated solutes in reducing the pressure field, or the solutes' screening effect.  \cite{39} examined solute segregation near dislocation and showed that the CE model overestimates the segregation, but the Larché-Cahn (LC) model \citep{40,41,42,43}, which accounts for a screening effect, gives a more accurate prediction.  Based on the LC model, the solute concentration profile can be expressed as
\begin{equation}
\label{Eq:7}
    c(r)=c_{0} - 3\omega \alpha \left(\frac{\partial M }{\partial c}\right) _{0}^{-1} P,
\end{equation}
where $\omega$ is the atomic volume of the matrix; $\alpha$ is the compositional strain coefficient; and $M$ is the diffusion potential of the matrix, which is the derivative of the Gibbs free energy with respect to solute concentration.  The subscript 0 in $(\frac{\partial M}{\partial c})_{0}^{-1}$ indicates the value of $(\frac{\partial M}{\partial c})^{-1}$ in the stress-free solid solution far away from dislocations and other sources of stress.  The values of $\omega$, $a$, and $M$ are all calculated at an average concentration, $c_0$.  The pressure field, $P$, in the atmosphere, and used in \cref{Eq:7} is given by
\begin{equation}
    P = \frac{G b(1+\nu) sin \theta}{3 \pi [1 - \nu + 4(1 + \nu ) G_m \xi ] r},
\end{equation}
where
\begin{equation}
 \xi = \alpha ^{2} \omega \left(\frac{\partial M}{\partial c}\right)_{0}^{-1}.
\end{equation}

Dislocations also interact with precipitate particles by contributing additional terms in the formation energy
\begin{equation}
    \Delta G=V (\Delta g_v + \Delta g_s ) + A \sigma + w_1 + w_2,
    \label{Eq:Gwrt}
\end{equation}
where $w_1$ and $w_2$ account for dislocation-precipitate interaction energies.  $w_1$ represents the interaction of the dislocation's stress field with the strain field of the nucleus, and $w_2$ the so-called modulus effect, in which, for example, a region of higher elastic modulus increases the dislocation’s self-energy \citep{christian1975textbook}. The analytical form for $w_1$ is available for a spherical precipitate nucleus.  According to %Dollins and Barnett 
\cite{45,48},
\begin{equation}
    w_1 = V \frac{G_m^m b \epsilon}{\pi}\frac{9 K^p}{3 K^p + 4 G_m^m} \frac{sin \theta}{r} = \Lambda w_1,
\end{equation}
where $K$ is the bulk modulus; $b$ is the Burger’s vector, and superscripts $m$ and $p$ represent the matrix and the precipitate, respectively.  Variables $\theta$ and $r$ are the cylindrical coordinates of the center of the precipitate nucleus on the plane normal to the edge dislocation.  The origin of the plane is located at the dislocation core. The $w_2$ term becomes significant when the shear modulus difference is considerable, like the order of magnitude difference in the Nb-N system \citep{45}.  In the binary Mg-Al system, its contribution is small because the shear moduli of the matrix and the precipitate, 18 and 25 GPa respectively, are reasonably close \citep{47,46}.\\

Assuming a spherical shape, the critical nucleus radius $R^*$ and the nucleation barrier $\Delta G^*$ can be obtained by minimizing $\Delta G$ in \cref{Eq:Gwrt} 
%I have to change rho since that's assigned to dislocation density. Changed to radius r
with respect to the nucleus radius, $R$.  In this case, we ignore $w_2$ so $R^*$ and $\Delta G^*$ can be solved as 
\begin{equation}
    R^* = \frac{-2\sigma}{(\Delta g_v+ \Delta g_s + w_1)}
\end{equation}
and
\begin{equation}
    \Delta G^* = \frac{16 \pi \sigma^3}{3} (\Delta g_v +\Delta g_s + w_1)^2.
\end{equation}
All energy terms above are functions of the solute concentration and are affected by solute segregation around an edge dislocation.  A comparison of these terms' magnitudes, analyzed in \cite{ma2019dynamic}, shows that the dominating terms are the driving force and the interfacial free energy, as shown in \cref{fig:SoluteDislocationSegregation}(b) and (c). The barrier can be effectively reduced by several orders of magnitude.  Although the number of the heterogeneous nucleation sites near dislocations is much lower than the number of the homogeneous nucleation sites throughout the whole system, this difference is still outweighed by the energy barrier reduction that arises due to heterogeneous nucleation.  For these reasons, solute segregation is a substantial factor in facilitating enhanced precipitate nucleation, and we argue that it is the primary thermodynamic factor when dislocations are serving as heterogeneous nucleation sites.  However, other mechanisms likely play a role as well, and these are discussed next.\\
\footnotetext[7]{Reprinted from Acta Materialia, Vol 172, X.L. Ma, Suhas Eswarappa Prameela, Peng Yi, Matthew Fernandez, Nicholas M. Krywopusk, Laszlo J. Kecskes, Tomoko Sano, Michael L. Falk, Timothy P. Weihs, Dynamic precipitation and recrystallization in Mg-9wt.\%Al during equal-channel angular extrusion: A comparative study to conventional aging, Pages 185-199, Copyright 2019, with permission from Elsevier.}

\subsubsection{Mechanisms of vacancy induced clustering and nucleation}
\label{vacancyhetnucleation}
\vspace*{1mm} 

Vacancies are another type of defect produced extensively during deformation that may facilitate precipitate nucleation.  Although studies of vacancies in Mg are rare, it was demonstrated that the vacancy concentration in Mg alloys under severe deformation can be as high as those of other metals, reaching concentrations of $10^{-4}$, as observed in experiments \citep{Sun2019,50} and MD studies. These high concentrations are due to the out-of-plane motion of non-basal dislocations in Mg \citep{yi2017solute, YIclimb2021}.  Vacancy clusters with sizes around 100 lattice sites in deformed samples were observed using positron annihilation spectroscopy (PAS) \citep{wu2018mechanistic}. Therefore, we discuss some related work on vacancy-facilitated precipitation in other metal systems to inspire our examination of Mg alloys.\\

\begin{figure}[h]
    \centering
    \includegraphics[width=\linewidth]{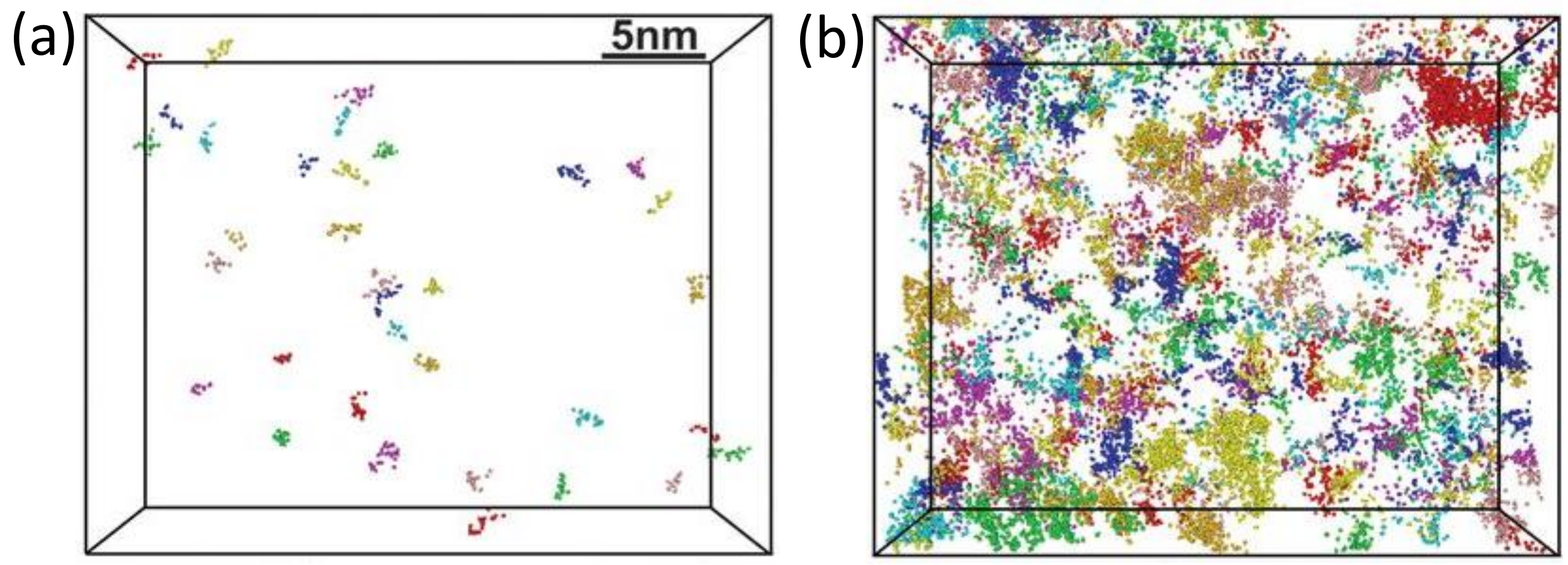}
    \caption{Atom probe reconstructions showing Mg-Cu clusters in an AA2024 alloy containing nine or more solute atoms in samples with identical thermal histories, without (a) and with (b) the cyclic deformation treatment, respectively. Individual clusters are colored differently. From \citep{Sun2019}. Reprinted with permission from AAAS.}
    \label{fig:cyclicdeformation}
\end{figure}

\cite{Sun2019} have shown that room temperature cyclic deformation of solutionized Al alloys increases their strength substantially, to values similar to peak-aged materials. The increases are attributed to the formation of a high density of fine (a few nm) solute clusters, as shown in \cref{fig:cyclicdeformation}, assisted by the vacancies generated by dislocation jogs, which are produced by deformation. The explanation proposed was that, due to vacancy-solute binding, the vacancy-assisted solute diffusion leads to the high number of solute clusters.  The binding between vacancy and solutes was also found to stabilize the solute clusters in Al-Mg-Cu alloys \citep{ZHU2021116455}.  This binding can even be utilized to reserve and release vacancies for the control of precipitation aging \citep{Pogatscher2014}.\\

It is worth noting that the formation of such fine solute clusters as those reported in \cite{Sun2019} was also common in two other distinct processes: aging of solid solutions fast-quenched from high temperatures and aging of irradiated materials \citep{Porter-Easterling, Ardell2016}, both of which are capable of retaining or producing a large number of vacancies. In the past, due to limitations in the spatial resolution of experimental characterization, these fine solute clusters or precipitates could be directly observed only near dislocation loops, which form from the collapse of vacancies, but could not be observed at vacancies or vacancy clusters. \cite{Sun2019}, however, were able to visualize solute clusters near vacancies using HAADF-STEM, providing direct evidence.\\
% https://www.overleaf.com/project/5f80747257d3740001250292

Besides enhancing the formation of solute clusters and precipitates kinetically by enhancing diffusion, vacancies can also thermodynamically promote the formation of solute clusters and precipitates by acting as heterogeneous nucleation sites when they agglomerate and form vacancy clusters and voids.  Examples of such aggregation into voids were first reported in radiated materials \citep{JohnsonLam1976, Okamoto1979} about four decades ago.
Large voids in irradiated Mg were observed more recently using high-resolution TEM \citep{Xu2013}. APT methods were also successfully used to probe the composition profile near voids in irradiated Ni-based materials~\citep{Wang2020}.  Efforts to understand and simulate solute cluster and precipitate formation near voids are ongoing.\\

\subsubsection{Density of dislocations and vacancies}
\vspace*{1mm} 

\begin{figure}[h]
    \centering
    \includegraphics[width=\linewidth]{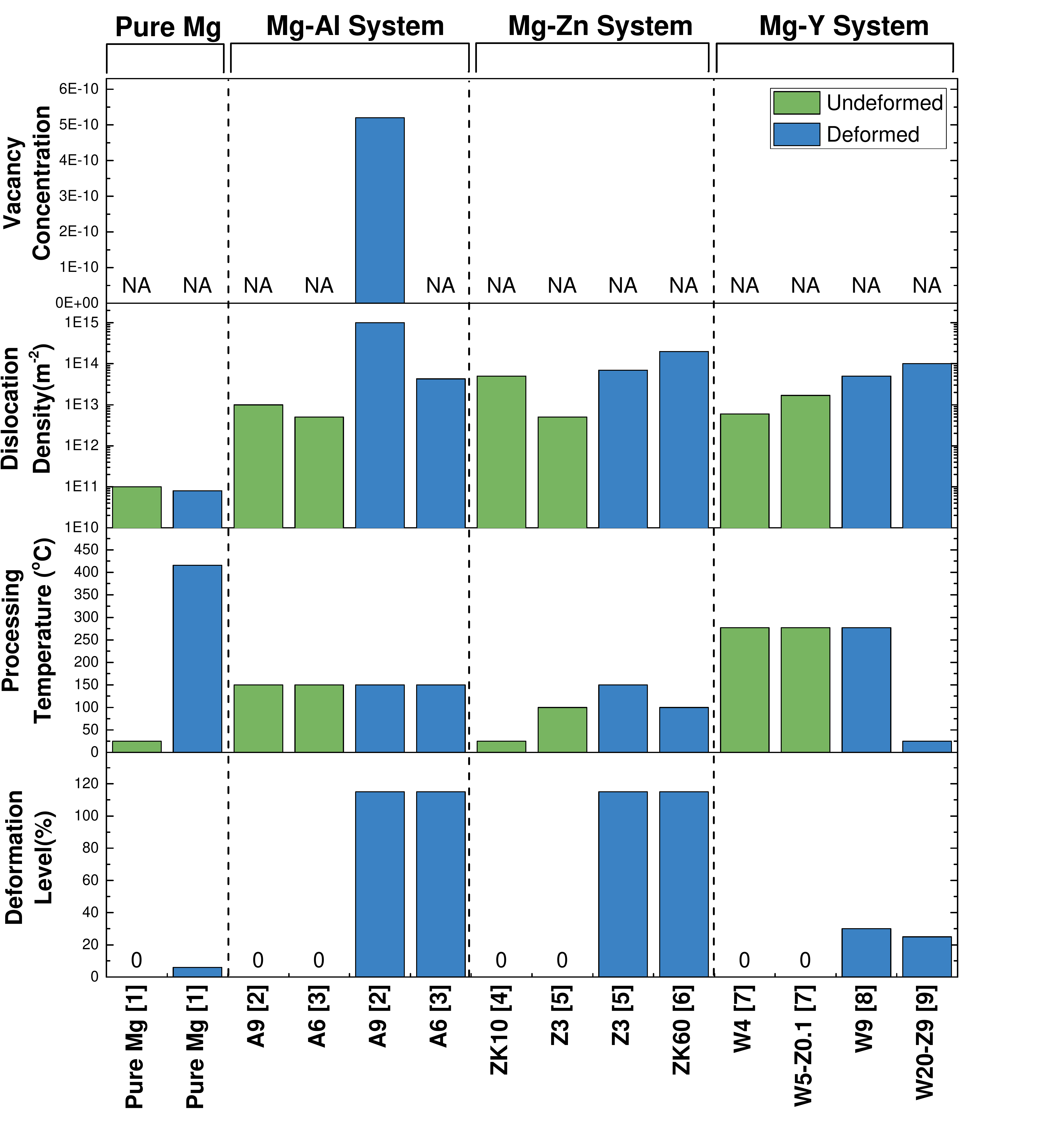}
    \caption{Statistics on concentration of defects and vacancies in deformed and undeformed pure Mg, Mg-Al, Mg-Zn, and Mg-Y alloy systems. Data from sources as indicated: 
    [1]\cite{lambri2007role}, [2]\cite{ma2019dynamic}, [3]\cite{prameela2020deformation}, [4]\cite{ha2019influence}, [5]\cite{prameela2020enhanced},  [6]\cite{balogh2010contributions}, [7]\cite{suzuki2003strengthening}, [8]\cite{maruyama2002creep}, [9]\cite{defectref7}.
    Undeformed samples are indicated with deformation levels labeled as `0', and unavailable data points as `NA'.}
    \label{fig:DefectDensity}
\end{figure}

As we discussed above, solute clusters and precipitates predominantly form around  heterogeneous nucleation sites like dislocations and vacancy clusters.  Therefore, the number density of these heterogeneous nucleation sites, which controls the spacing between solute clusters and precipitates, becomes an important tuning parameter during processing.  While equilibrium densities of dislocations and vacancies are relatively well-known (the number densities of dislocations can vary from $10^{10}-10^{12}$ m$^{-2}$ and the equilibrium vacancy concentration is about $10^{-6}$ \citep{Porter-Easterling}), these can both vary significantly during deformation. Experimentally, dislocation density can be estimated by TEM or XRD, and vacancy concentration can be determined by XRD. \cref{fig:DefectDensity} shows the values of dislocation density and vacancy concentration for pure Mg and alloys belonging to the Mg-Al, Mg-Zn, and Mg-Y systems. For each category, both deformed and undeformed cases are reported. Deformation was accomplished through a range of processes, such as compression, ECAE, and torsion. From these results, it is evident that the dislocation density and vacancy concentration are much higher in deformed samples than in undeformed samples.\\

Strain hardening models offer a direct connection between the number densities of dislocations and/or twins and the strain.  A review of dislocation-based strain hardening models can be found elsewhere \citep{24}; also, more recent developments take into consideration the activity of twinning \citep{25}.  \\

In general, changes in the flow stress depend on the dislocation structure and are related to the dislocation density, $\rho$, as $\sigma = \alpha MG_mb\rho^{1/2}$, where $b$ is the Burgers vector, $G_m$ is the shear modulus of Mg, $\alpha$ is a numerical constant assumed to be 1.0\citep{RN1925}, and $M$ is the Taylor factor which assumes the value of 4.5 \citep{RN1925}.  Based on the Luk\'{a}\u{c}-Bal\'{i}k model \citep{lukac1995kinetics}, the dislocation density, $\rho$, evolves as a function of strain as
\begin{equation} \label{eq:27}
    \frac{d\rho}{d\epsilon}=K_1+K_2\rho^{1/2}-K_3\rho-K_4\sigma\rho^{3/2},
\end{equation}
where the first term is dislocation nucleation and multiplication due to impenetrable obstacles (grain boundaries, incoherent precipitates, dispersion particles) and is proportional to the density of these obstacles. This term is normally assumed to be a constant during the processing, but in reality, both the grain and precipitate microstructures will change.  The second term is dislocation multiplication by forest dislocations, and the third and fourth terms are dynamic recovery (i.e. dislocation annihilation due to cross-slip) and climb, respectively.  The other two models mentioned above only include two or three of the terms used in the Luk\'{a}\u{c}-Bal\'{i}k model.  Correspondingly, the work hardening rate is
\begin{equation} \label{eq:28}
    \Theta \equiv \frac{d\sigma}{d\epsilon} = \frac{A}{\sigma-\sigma_y} +B -C(\sigma-\sigma_y) -D(\sigma-\sigma_y)^3,
\end{equation}
where $\sigma_y$ is the yield stress, and parameters $A$, $B$, $C$, and $D$ are proportional to $K_i$ ($i$=1,2,3,4), respectively:
\begin{equation}
\begin{split}
    A & = M^3(\alpha Gb)^2(\frac{\dot{\epsilon}}{\dot{\epsilon_0}})^{2/n} \frac{K_1}{2},\\
% \end{equation}
% \begin{equation}
    B & =M^2\alpha Gb(\frac{\dot{\epsilon}}{\dot{\epsilon_0}})^{1/n}\frac{K_2}{2},\\
% \end{equation}
% \begin{equation}
    C&=M\frac{K_3}{2},\\
% \end{equation}
% \begin{equation}
    D& =(\frac{\dot{\epsilon}}{\dot{\epsilon_0}})^{-1/n} \frac{K_4}{2M^2 \alpha Gb}.
\end{split}
\end{equation}

\begin{figure}[h]
    \centering
    \includegraphics[width=\linewidth]{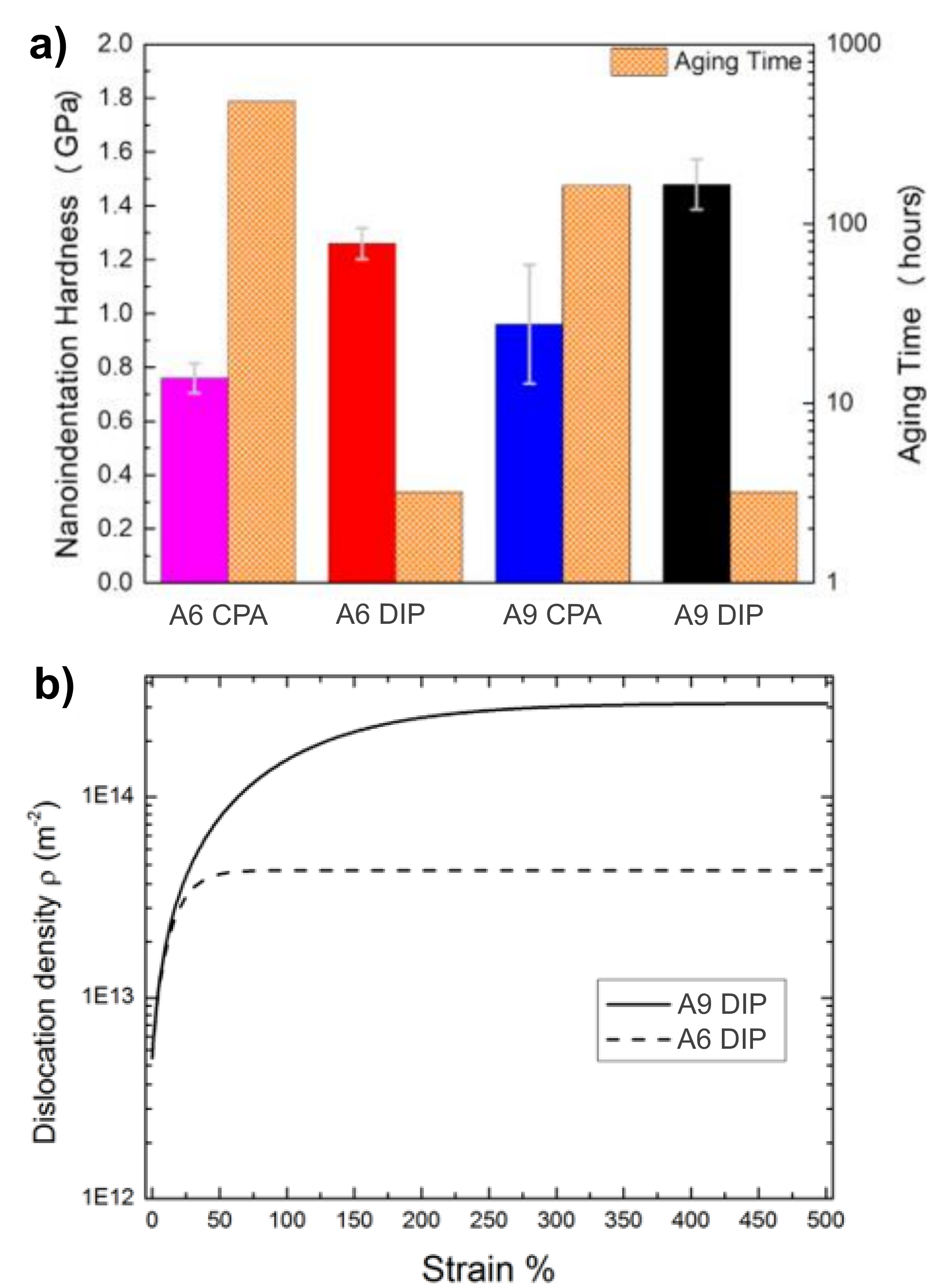}
    \caption{(a) Aging times and nanoindentation hardness in grain interiors of A6 and A9 samples following two different processing histories at 150 $^{\circ}$C: conventional peak-aged (CPA) and deformation induced precipitation (DIP); (b) Dislocation density as a function of strain for A6 DIP and A9 DIP samples, predicted based on the Lukác-Balík model, see text for details. This figure is taken from \cite{prameela2020deformation}.}
    \label{fig:strainhardening}
\end{figure}

A recent study by \cite{prameela2020deformation,21} obtained parameters $A,B,C,$ and $D$ by fitting \cref{eq:28} to experimental data for Mg-Al alloys with various Al content.  Then, authors predicted the evolution of dislocation density for ECAE processing of two Mg-Al alloys, A6 and A9. Strain rate was chosen to be $10^{-4} s^{-1}$ and $\epsilon_0$ was used as normalization parameter.  The stress exponent $n$ is on the order of $10^2$ \citep{lukac1995kinetics,RN1926}, so $(\dot{\epsilon}/\dot{\epsilon_0})^{-1/n}$ can be approximated as 1.0.\\

The resulting evolution of dislocation density is shown in \cref{fig:strainhardening}(b) as a function of strain. The difference in dislocation density generated during processing correlates with differences in precipitate density for A6 and A9 alloys.  Similar results were obtained for Mg-Zn alloys \citep{prameela2020enhanced}, strongly suggesting that the main enhancement is due to dislocations forming through deformation.  In these studies, twinning was not significant.  Normally, twinning in Mg systems is much more active than in the Al systems due to the smaller number of slip systems.  Integrating the twin fraction into the strain hardening model could be important when considering twin-induced precipitate nucleation.  In addition, the strain hardening model applied here is a phenomenological model where the state of micro-structure is contracted to one parameter, density of dislocations \citep{Kocks2003}.  More sophisticated crystal plasticity models can be used to consider the state variables' spatial and orientation distributions.\\

Vacancy evolution is more complex to model because both the generation and annihilation of vacancies depend on other defects like dislocations. The difficulty of observing vacancies and vacancy clusters in experiments adds an additional challenge.  \cite{Militzer1994,Fischer2011} have proposed continuum models for the generation and annihilation of vacancies during deformation.  In terms of vacancy clustering, quantitative models are under development to predict the lifetime and size distribution of vacancy clusters/voids by MD simulation \citep{52}, Kinetic Monte Carlo simulation (kMC) \citep{53}, phase filed modeling \citep{54}, and continuum modeling \citep{55,56}. The mobility of vacancy clusters/voids could be much lower than that of vacancies, allowing them to participate in the precipitate nucleation process for a much longer time.  This is consistent with the observation of vacancy clusters/voids, as well as the solute segregation near them, as mentioned previously in \cref{vacancyhetnucleation}.\\ 

Single vacancies have a short lifetime, as they diffuse to sinks like grain boundaries quickly.  Their lifetimes have often been studied using the PAS method, which has long been applied to probe lattice defects, including vacancies and small voids in metallic materials \citep{vcivzek2018characterization}.  The average lifetime of a positron, $\bar{\tau}$, is a function of the density and size of vacancies and voids.  For example, the lifetime of positrons interacting with a perfect Mg structure is 225 ps, and that of positrons interacting with monovacancies in Mg structure is 255 ps \citep{hautojarvi1982trapping}. The concentration and size of vacancies and voids can be estimated by observing the average lifetime of the positrons.  The positron lifetime for all elements can be found in \cite{robles2007positron, zhang2015accurate}.
Early experimental work shows low vacancy densities in Mg \citep{hautojarvi1982trapping}. This is because dislocations in Mg are more uniformly distributed, unlike in bcc or fcc metals, which form cellular sub-structures and are highly non-uniform.  The more uniform dislocation sub-structure helps vacancies escape \citep{vcivzek2019development}. However, later experiments show different results \citep{hautojarvi1982trapping, abdelrahman1996positron, lambri2007role}.   It is possible that severely deformed Mg materials may generate a cellular sub-structure, as indicated by experiments \citep{lee2018dynamic,li2008dislocation}, creating zones that are nearly dislocation-free, resulting in a longer lifetime for vacancies and vacancy clusters that may enhance interplay with solutes.  Future experiments are planned to explore the density of vacancies and vacancy clusters in these model binary Mg alloys, and in Mg-Zn alloys in particular.\\

\subsection{Precipitation modeling}
\label{subsec:Precipitationmodeling}
Various models have been proposed to predict the formation of precipitates in metallic alloys, many of which are based on CNT \citep{aaronson1999kinetic} and the Lifshitz-Slyozov-Wagner theory (LSW)  \citep{lifshitz1961kinetics, wagner1961theorie}. Models based on classical nucleation theory include the works of \cite{myhr2000modelling}, \cite{paliwal2019precipitation}, and \cite{hutchinson2005modeling}. \cite{svoboda2004modelling} described the nucleation mechanism of a multi-phase, multi-component alloy, yet computed the evolution of each precipitate individually and considered spherical precipitates only, both of which render this approach computationally unfeasible for larger problems. \cite{vaithyanathan2004multiscale} used a physically informed multiscale model to compute the formation and coarsening of precipitates in as-cast Al-Cu alloys.\\

\cite{perez2008implementation} provided a comparison of three different approaches, including classical nucleation and growth theory (CNGT), viz.\ the ``mean-radius approach", the Euler-like,  and the Lagrange-like multi-class approaches. 
The mean-radius approach presents a fairly simple approximation of the evolution of precipitates in a solid. It is assumed that nucleation and growth of the nuclei occur concurrently. The evolution of the mean precipitate radius is thus given by an evolution law based on LSW theory:
\begin{equation}
\frac{d\bar R}{dt}|_{\text{growth}} = \frac{D}{\bar{R}} \frac{X - X^{i}(\bar{R} )}{\alpha X^{P} - X^{i}(\bar{R}} + \frac{1}{N} \frac{dN}{dt}(R^*_{k_{\text{B}}T} - \bar{R}),
\end{equation}
where $\bar R$ stands for the mean radius, $D$ is the diffusion coefficient, $X^P$ the molar composition of the precipitate (which is often assumed to be stoichiometric and constant), $X$ is the equilibrium solute molar fraction in the matrix, and $X^{i}$ is the equilibrium molar fraction at the precipitate/matrix interface. The conditions for the validity of the LSW treatment of coarsening are strict; notably, there is no elastic interaction, diffusion is seen as the rate-limiting process, and the volume fraction of particles is negligible. \cite{voorhees1992ostwald} found that a non-negligible precipitate volume fraction has an accelerating effect on the coarsening kinetics and can be accounted for by a constant, multiplicative correction factor that depends solely on the volume fraction. \\

\cite{hutchinson2005modeling} modeled precipitation in the Mg alloy AZ91. Using classical nucleation theory, they laid out an improved description of growing non-spherical precipitates, as is observed in AZ91. This study focused on Mg$_{17}$Al$_{12}$ ($\beta-$phase) precipitates \textemdash{}, which form as nuclei of spherical shape and evolve into laths, assumed to be of constant length-thickness, length-height, and height-thickness ratios, lying in basal planes. The precipitation process is divided into phases: a nucleation and initial growth phase, where both processes are treated concurrently, and a coarsening phase, which begins when an equilibrium fraction of precipitates in the solute is reached. Diffusional growth theory is applied to compute the evolution of the lath length, using the Zener-Hillert treatment of plate lengthening \citep{hillert1957pressure, hillert2003diffusion}. At small sizes, particles are assumed to form as spheres which then evolve into laths; this assumption is the basis for their derivations. The coarsening of a spherical precipitate of equivalent volume as the lath is computed, and the new length of the precipitate is obtained using the updated radius (hence approximated by the LSW relations).\\

A complete picture of precipitation also necessitates accounting for the interplay of slip and twinning with precipitates after their formation. To this end, \cite{RN2118} modeled the interaction of slip and, especially, tensile twins with preexisting periodic arrays of precipitates, showing how the precipitate concentration, size, shape, and orientation affect the motion of passing twin boundaries. They demonstrated the effects of precipitate spacing, size, and orientation on the strain hardening due to pinning interactions with migrating twin boundaries. Although the study was limited to a single twin variant and simple precipitate configurations, it gave interesting insight into the microstructural mechanisms leading to hardening by precipitates in the presence of slip and twinning in Mg.\\

% The effect of stress on the nucleation and growth of precipitates in Al alloys was investigated by \cite{skrotzki1996effect}. Their results hint at a possible correlation of the orientation of the precipitates and the stress; however, the behaviour in Al alloys does not allow an equivalent statement for Mg alloys. Nevertheless, their results show that the precipitates prefer to align with the applied stresses, given it exceeds a ceratin critical value. Further, this observation is in part generally valid, since multiple precipitates show the same tendency but for different stress levels. 
% In addition to the increased propensity for precipitates to align with an applied stress, it was noted that the density of $\theta'$ precipitates increases with rising stresses along certain crystallographic planes.

% In a crystal plasticity setting, it is still tricky to find a good description of how the local stress state influences the formation of new precipitates. The hardening effects of precipitates on the other hand are pretty well understood and can relatively easily be incorporated. Efforts in our future work will among other resolve around obtaining a sound formulation for the combination of crystal plasticity and nucleation models. Information about the link between the plastic state and the nucleation laws are therefore required. Ideally this can be obtained from lower scales simulations. 

Our current work focuses on including a precipitation model within the crystal plasticity framework on the mesoscale in a two-way coupled fashion: the local mechanical state of deformation affects precipitation, while precipitates impact mechanical hardening through the effects discussed in \cref{sec:EffectSolute} and \cref{sec:ControllingClustering}. Nucleation, growth and coarsening of precipitates, as described by \cite{hutchinson2005modeling}, can be used to compute the steady-state nucleation of new precipitates on the fly and thus (in part) close the loop between the different effects: nucleation of precipitates, solute diffusion, and (local) hardening. A key open challenge is the lack of detailed insight into the influence of the local stress state and microstructure on precipitate nucleation and growth. The intricate interactions of lattice defects (dislocations and voids generated by plastic deformation), twins (abundant in Mg), precipitates, and possibly mechanisms of recrystallization at elevated temperature make the creation of a predictive scale-bridging model (from nano-precipitates to micron-scale grains) an open challenge.

\section{Enabling inverse design of metal alloys}

\begin{figure*}[h!]
    \centering
    \includegraphics[width=0.85\linewidth]{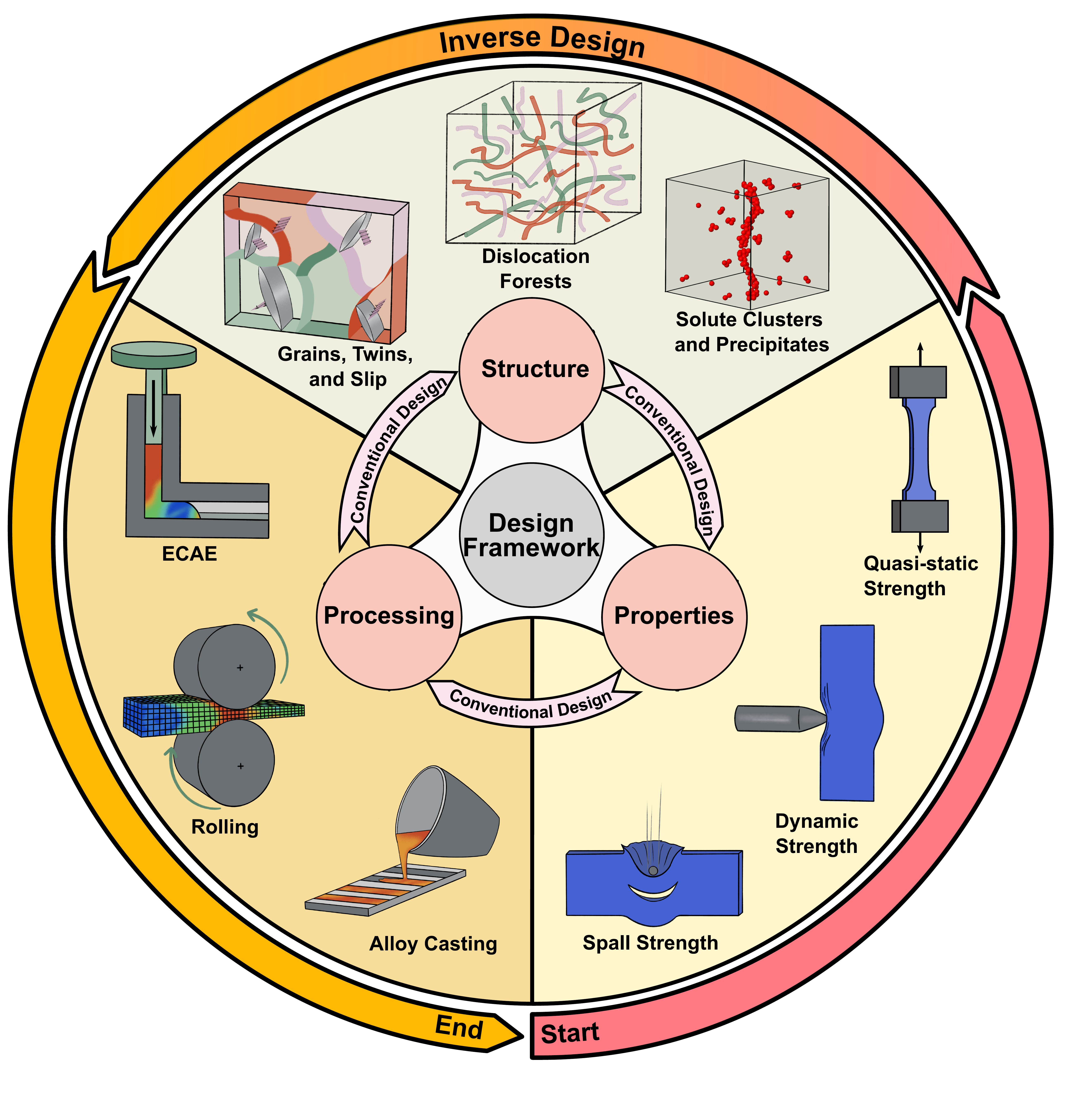}
    \caption{Schematic illustrating the conventional and inverse design of Mg alloys through experiments and simulations. In the conventional design process, different processing methods (such as casting different compositions, and thermomechanical processing treatments) to alloys produces different structural features in the alloy such as grain and twin size and properties, dislocation densities, and solute cluster or precipitate formation, which in turn result in measurable properties, which can inform future iterations of processing to achieve improved properties. In the inverse design method, desired properties determine the target microstructure of the alloy, which in turn can be used to design or select an appropriate processing method.}
    \label{fig:wheel}
\end{figure*}

%Throughout this paper we have identified key mechanisms operating on the micro- and mesoscale, which govern the macroscopic material properties of Mg alloys. In particular, we have shown how the micro- and nano-scale mechanisms -- solute clustering and precipitation-- affect the mesoscale slip and twin evolution (cf.~\cref{sec:EffectSolute}) and ultimately the macroscopic material behavior. We have further reviewed processing techniques that are capable of controlling clustering and nucleation processes (cf.~\cref{sec:ControllingClustering}) as well as modeling approaches capable of describing some aspects of the underlying physics (cf.~\cref{sec:ModelingApproaches}), albeit in isolation from each other. 

The findings and the corresponding data presented in the above described studies contribute to an understanding of process-structure-property relationships required to lay the groundwork for a more robust ICMD approach. In \cref{sec:Introduction}, we briefly discussed the origin of this approach and some recent applications of it to Mg alloy design. At the heart of this design cycle is the canonical relationship between processing, structure and properties illustrated in  \cref{fig:wheel}. Conventional materials design seeks to characterize the microstructure resulting from various processing methodologies, and, by correlating the observed microstructures with the relevant properties, to enable iterative feedback so that the processing may be adjusted to systematically and incrementally improve  properties.
%As discussed previously, the macroscopic behavior of Mg alloys is complex and results from a multitude of mechanisms operating over a broad range of disparate scales, posing a significant challenge. Therefore, compared to the conventional forward approach, the inverse design approach (\cref{fig:wheel}) is challenging.\\ 
By contrast, the ICMD framework seeks to move us closer to an inverse approach. Instead of long and costly iterations towards an improved processing path, we aim to consider materials design as an optimization problem. Target properties would be used to specify microstructures of interest, and both chemistries and processes would be developed specifically to achieve such microstructures \citep{mallick2019shock, prameela2020defect, mallick2019role}. This approach is anticipated to save the cost of experimental trial-and-error across the vast available processing parameter space. \\

The least developed aspect of this grand vision has been the processing-structure connection. In many respects, \cref{fig:wheel} does not do justice to the complexity of the interrelationship between structure and processing. It is our hope that, 
by mapping out a small portion of the complex interplay between structure and processing in the above sections and by reviewing the present state of knowledge about the underlying physics, we identify a way forward towards a more robust ICMD approach. To support ICMD inverse design, we must generate knowledge of how processing induces changes in microstructure that, in turn, alter processing outcomes. We strongly believe that closely coupled experimental and computational studies as envisioned within the ICME framework \citep{national2008integrated} are required to generate this knowledge base.\\

Clearly, this integrated material-by-design strategy for Mg alloys will require improved accuracy and completeness in the data, obtained both by experiments and simulations, upon which we base our models of the processing-structure interrelationship. While we have revealed some of the governing mechanisms of this interrelationship and revisited the modeling techniques for individual mechanisms (e.g., MD for vacancy and solute clustering, crystal plasticity for slip and twinning, nucleation theory for precipitation), the dramatic separation of these mechanisms in spatial and temporal scales still poses a serious challenge.  \\

\section{Conclusions and future directions}

We have discussed three main strengthening mechanisms in Mg alloys: solid solution strengthening, solute cluster strengthening, and precipitation strengthening. In particular, we have showcased how these mechanisms manifest in Mg-Al, Mg-Zn, and Mg-Y alloys. We have further highlighted how to manipulate the microstructure of Mg alloys using thermomechanical processing to improve mechanical properties. The main conclusions of the paper are the following: 

\begin{enumerate}

   \item 	Dissimilar slip systems and hexagonal crystal symmetry make the design of Mg alloys for structural applications a challenging exercise.  To overcome this, we propose focusing on structure-property and processing-structure relationships for three Mg alloy systems (Mg-Al, Mg-Zn, Mg-Y). 

   \item 	For structure-property relationships, it is important to understand the interaction of solute atoms, solute clusters, and precipitates with dislocations, twins, grain boundaries, and texture. Each of these interactions is strongly influenced by solute chemistry, solute concentration, temperature, and loading direction relative to the slip and twinning planes. 

   \item 	For processing-structure relationships, we have shown that homogeneous nucleation is insufficient to generate rich solute clusters or high densities of nanoscale precipitates. We showcased a new processing strategy based on heterogeneous nucleation called DIP (Deformation Induced Precipitation), through which one can inject atomic-scale defects such as dislocations and vacancies at low temperatures to boost nucleation kinetics and produce fine precipitates. Furthermore, we have discussed approaches to quantify the defect density evolution in Mg alloys through experiments and modeling. 

   \item 	 We have highlighted relevant modeling approaches to describe slip, twinning, and other deformation mechanisms during thermo-mechanical processing conditions. In addition, we have also discussed how these deformation events can influence solute clustering, nucleation and growth of precipitates. These provide valuable information to atomistic and mesoscale modeling frameworks such as MC, MD and DD to unravel interactions of solute atoms, point and line defects, and solute cluster and precipitate formation. 

   \item 	We have outlined an integrated material-by-design approach, which combines the ``structure-property" and ``processing-property” approaches into a unified design scheme. We argue that inverse design can be facilitated by high-throughput data collection and a more robust ICMD approach.
   
\end{enumerate}

Robust and predictive frameworks are critical for understanding the complex microstructural evolution during thermomechanical processing. Through this article, we have aimed to convey how modeling and experimental approaches can be used to design advanced Mg alloys by maximizing the alloys' strengthening by solute atoms, solute clusters, and nanoscale precipitates.
%solution hardening and precipitation strengthening.

\section*{Conflict of interest statement}
The authors declare no conflict of interest.

%\section {Supplementary Materials}

%\href{https://www.overleaf.com/2297443229qbfmtbyftxcs}{link}

\section*{Acknowledgements}
The authors would like to gratefully acknowledge the financial and technical support from Center for Materials under Extreme Dynamic Environment (CMEDE). Research was sponsored by the U.S. Army Research Laboratory and was accomplished under Cooperative Agreement Number W911NF-12-2-0022. The views and conclusions contained in this document are those of the authors and should not be interpreted as representing the official policies, either expressed or implied, of the U.S. Army Research Laboratory or the U.S. Government. The U.S. Government is authorized to reproduce and distribute reprints for Government purposes notwithstanding any copyright notation herein.

\bibliography{mybibfile}

\end{document}